\documentclass[12pt]{article}

\usepackage{microtype} 

\usepackage[numbers,sort&compress]{natbib}
\usepackage{graphicx}
\usepackage{amsmath}%
\usepackage{mathtools} 
\usepackage{amsthm}
\usepackage{amsfonts}
\usepackage{amssymb}
\usepackage{bbm}
\usepackage{fullpage} 
\usepackage{color}
\usepackage{latexsym}
\usepackage{subfigure}
\usepackage{soul}
\usepackage{enumerate}
\usepackage{multirow}

\usepackage{times}
\usepackage[utf8]{inputenc}
\usepackage[T1]{fontenc}
\usepackage[nameinlink]{cleveref}
\usepackage{thm-restate}
\usepackage{xfrac}

\theoremstyle{plain}
\newtheorem{thm}{\protect\theoremname}[section]
  \theoremstyle{plain}
  \newtheorem{prop}[thm]{\protect\propositionname}
 \newtheorem{example}[thm]{Example}\newtheorem{definition}[thm]{Definition}\theoremstyle{plain}
 \newtheorem{rem}[thm]{\protect\remarkname}
  \theoremstyle{plain}
  \newtheorem{lem}[thm]{\protect\lemmaname}
  
  \providecommand{\lemmaname}{Lemma}
  \providecommand{\propositionname}{Proposition}
 \providecommand{\remarkname}{Remark}
\providecommand{\theoremname}{Theorem}

\usepackage[ruled,vlined, norelsize]{algorithm2e}

\newcommand{\old}[1]{{{}}}

\graphicspath{{./Figures/}}

\def\D{\mathcal{D}}
\def\L{\mathcal{L}}

\def\I{\mathcal{I}}

\def\Po{\rm{Po}}

\def\K{\bf{K}}

\usepackage{xspace}

\usepackage[margin=1in]{geometry}

\def\Thanks#1{\gdef\thefootnote{\arabic{footnote}}\thanks{#1}}
\def\ThanksComma#1{\gdef\thefootnote{\arabic{footnote},}\thanks{#1}{
}}

\usepackage{diagbox}
\usepackage{makecell}
\usepackage{array,booktabs}
\usepackage{hhline}
\usepackage[table]{xcolor}

\usepackage{lineno}

\begin{document}
\title{A Model of Random Industrial SAT}

\newenvironment{Proof}[1]{\par\noindent{\bf Proof{#1}:}\quad}{} 

\author{D. Barak-Pelleg\Thanks{
Department of Mathematics, Ben-Gurion
University, Beer Sheva 84105, Israel.
E-mail: dinabar@post.bgu.ac.il}
 \and  D.~Berend\ThanksComma{
Departments of Mathematics and Computer Science, Ben-Gurion
University, Beer Sheva 84105, Israel.
E-mail: berend@math.bgu.ac.il}
\Thanks{Research supported in part by the Milken
Families Foundation Chair in Mathematics.}
\and
J.C.~Saunders\ThanksComma{
Department of Mathematics, Ben-Gurion
University, Beer Sheva 84105, Israel. E-mail: saunders@post.bgu.ac.il}
\ThanksComma{Research supported by an Azrieli Fellowship from the Azrieli Foundation.}\hspace{4pt}\Thanks{Current address:
Department of Mathematics \& Statistics,
University of Calgary, 
2500 University Drive NW,
Calgary, Alberta, Canada,
T2N 1N4.}}
\date{}
\maketitle
\begin{abstract}
One of the most studied models of SAT is random SAT. In this model, instances are composed from clauses  chosen uniformly
randomly and independently of each other. This model may be unsatisfactory in that it fails to describe various  features of SAT instances, arising in real-world applications. Various modifications have been suggested to define models of industrial SAT. Here, we focus mainly on the aspect of community structure. Namely, here the set of variables consists of a number of disjoint communities,
and clauses tend to consist of variables from the same community. Thus, we suggest a model of random industrial SAT, in which the central generalization with respect to random SAT is the additional community structure.

There has been a lot of work on the satisfiability threshold of random $k$-SAT, starting with the calculation of the threshold of $2$-SAT, up to the recent result that the threshold exists for sufficiently large $k$. 

In this paper, we endeavor to study the satisfiability threshold for the proposed model of random industrial SAT. Our main result is that the threshold in this model tends
to be smaller than its counterpart for random SAT. Moreover, under
some conditions, this threshold even vanishes.

\end{abstract}

\section{Introduction}

For both historical and practical reasons, the Boolean satisfiability problem (SAT) is one of the most important
problems in theoretical computer science. It was the
first problem proven to be NP-complete \citep{cook}. Since  its introduction, there has been growing interest in the problem, and many aspects of the problem have been researched. 

In this problem, one is required to determine whether a certain Boolean
formula is satisfiable. An instance of the problem consists of a Boolean
formula in several variables~$v_{1},\dots,v_{n}$. The formula is
usually given in conjunctive normal form (CNF). The basic building
block of the formula is a $\emph{literal}$, which is either a variable
$v_{j}$ or its negation $\overline{v}_{j}$. A $\emph{clause}$ is
a disjunction of the form~$l_{1}\lor...\lor l_{k}$ of several distinct
literals. Thus, altogether, the formula looks like $C_{1}\land\:C_{2}\land\:...\land\:C_{m}$,
where each $C_{i}$ is a clause, say~$C_{i}=l_{i,1}\lor...\lor l_{i,k_{i}}$.
Given a formula, one may assign a TRUE/FALSE value to each of the
variables $v_{1},\dots,v_{n}$. The formula is $\emph{satisfiable}$,
or SAT, if there exists an assignment under which the formula is TRUE,
and is $\emph{unsatisfiable}$, or UNSAT, otherwise. 

The $k$-satisfiability ($k$-SAT) problem is a special case of SAT,
in which each clause is a disjunction of up to~$k$ literals. Some
authors restrict $k$-SAT to instances with exactly $k$ literals
in each clause, which terminology we will follow here. Given $n$
and $k$, let~$\Omega\left(n,\:k\right)$ denote the set of all~$\tbinom nk 2^{k}$  possible clauses of length $k$ over $n$ Boolean variables. A $\emph{random}\:\emph{k-SAT}\:\emph{instance }$ with $m$ clauses
is a uniformly random element of~$\left(\Omega\left(n,\:k\right)\right)^m$.
Namely, it consists of $m$ clauses, selected uniformly randomly and independently from $\Omega\left(n,\:k\right)$.
Thus, clause repetitions are allowed, and two instances, differing in the order of the clauses only,
are considered as distinct.

The ratio $m/n$ is the $\textit{density}$ and denoted by $\alpha$.
This parameter turns out to be very important. If $\alpha$ is sufficiently
small, then a large random instance with density $\alpha$ is SAT
with high probability, whereas if it is sufficiently large
then a large random instance is UNSAT with high probability. Despite
its loose name, the notion of ``with high probability'' is well
defined. Let $\left(E_{j}\right)_{j=1}^{\infty}$ be a sequence of
events. The event~$E_{j}$ occurs $\emph{with high probability (w.h.p.)}$ 
if~$P\left(E_{j}\right)\xrightarrow[j\to\infty]{}1$. In our case,
we take larger and larger random instances with some fixed density,
and inquire whether they are SAT or UNSAT. For $k\ge2$, denote \citep{threshold achiloptas}:
\begin{align*}
r_{k} & \equiv\textmd{sup\ensuremath{\left\{  \alpha:\textmd{A random density-}\alpha\textmd{ instance is SAT w.h.p.}\right\} } ,}\\
r_{k}^{*} & \equiv\textmd{inf\ensuremath{\left\{  \alpha:\textmd{A random density-}\alpha\textmd{ instance is UNSAT w.h.p.}\right\} } . }
\end{align*}
For $k=2$, it was proved long ago \citep{threshold of 2 item ca,threshold of 2 item Fe,threshold of 2 item G1}
that~$r_{2}=r_{2}^{*}=1$. The Satisfiability Threshold Conjecture
claims that, in fact,~$r_{k}=r_{k}^{*}$ for all $k$ \citep{threshold of 2 item ca}.
This conjectured common value is the $\textit{satisfiability threshold}$.
It has been a subject of interest among researchers, theoretically
and empirically, to prove the conjecture for~$k\ge3$ and find the
threshold. Recently, the conjecture has been proved for large enough~$k$
\citep{large k}.

As part of this research, lower and upper bounds were obtained on
$r_{k}$ and $r_{k}^{*}$ for $k\ge3$. In \citep{threshold franco}
it was proven that~$r_{k}^{*}\le2^{k}\ln2$. This has been improved
in \citep{thresh kkk89} to~$r_{k}^{*}\le2^{k}\ln2-\frac{1}{2}(1+\ln2)+\varepsilon_k$.
From the other side, a sequence of successive improvements led finally to
the bound~$r_{k}\ge2^{k}\ln2-\frac{1}{2}(1+\ln2)+\varepsilon_k$ \citep{thresh coja 2016}.
Thus, with the satisfiability conjecture settled in \citep{large k}
for large~$k$, it follows that~$r_{k}=r_{k}^{*}=2^{k}\ln2-\frac{1}{2}(1+\ln2)+\varepsilon_k$
for such~$k$. For small values of~$k$, more specific results were
obtained. For~$k=3$, the best bounds seem to be~$r_{3}\ge3.52$
\citep{Sat3LowerBound2,Hajiaghayi}, 
and~$r_{3}^{*}\le4.4898$ \citep{Diaz}.
Experiments and other results of heuristics, based on statistical
physics considerations, indicate that~$r_{3}\approx4.26$ \citep{thresh zecchina 1,thresh zecchina 2},
$r_{4}\approx9.93$, $r_{5}\approx21.12$, $r_{6}\approx43.37$, $r_{7}\approx87.79$
\citep{thresh zecchina 1}. 

Much more is known about $2$-SAT. First, unlike~$k$-SAT for~$k\ge3$,
which is an NP-complete problem, $2$-SAT instances may be solved
by a linear time algorithm \citep{threshold of 2 item ca,threshold of 2 item G1}.
Also, there is quite precise information about $2$-SAT for density
very close to the threshold $r_{2}=1$ \citep{scaling window,Wilson results}. 

It has been argued that instances of random~$k$-SAT do not in fact
represent real-world, or industrial, instances \citep{Oh2015,Oh2016,Oh2018}.
One of the major differences between industrial and random SAT instances
is that the set of variables in industrial instances often consists
of a disjoint union of subsets, referred to as $\textit{communities}$;
clauses tend to comprise variables from the same community, with only
a minority of clauses containing variables from distinct communities~\citep{Jordy 2009C,newsham}.
There are several additional variations \citep{Jordy 2015,Jordy 2009B}. For example, the variables
may be selected non-uniformly (say, according to a power-law distribution
\citep{Jordy 2009 B1,Levy17}), and/or the clauses may be of non-constant
length.  

In this paper we work with a (generalization of a) model introduced
by \citep{Jordy 2015 modularity} . Our model is similar to the random
model, except for the partition of the variables into communities.
These communities are of the same size. There are several clause types (defined precisely in the next section),
differing in the number of variables from the same or distinct communities
in each clause. 

Our focus is on the satisfiability threshold in this model. The question
has been studied in \citep{Jordy 2015 modularity}, mostly experimentally,
for the model suggested there. We show that the findings in that paper,
whereby the threshold tends to be smaller when there are many single-community
clauses (i.e., clauses consisting of variables from the same random community) remain true in the general model. In fact, if the communities
are small, the threshold may even be $0$.

We present our model in Section \ref{newcom sec:new model}. The main
results are stated in Section 3, and the proofs follow in Section~4. In Section~5 we present some simulation results.

\section{\label{newcom sec:new model} A Model of Random Industrial SAT}


In industrial SAT, the strength of the community structure of an instance
is usually measured by its modularity \citep{Jordy 2009C,Jordy 2016,modularity newman B}.
Roughly speaking, given a graph, its modularity gives an indication
of the tendency of the vertices to be connected to other vertices,
which are similar to them in some way. In our case, an instance defines
the following undirected graph. The set of nodes is the set of variables~$\left\{ v_{1},\ldots,v_{n}\right\} $.
There is an edge~$(v_{i},v_{j})$ for~$i\ne j$ if there exists
a clause in the instance, containing both variables~$v_{i}$ (or
its negation) and~$v_{j}$ (or its negation). More precisely, we view this object as a multi-graph; if both $v_i$ and $v_j$ (or their negations) appear in several clauses, there are several edges connecting them. Given an instance,
high modularity indicates that there exists a partition of the set
of variables into subsets, such that a large portion of the edges
connect vertices of the same subset, compared to a random graph with
the same number of vertices and same degrees \citep{modularity newman A,modularity newman B}
. 

As in the regular model, we have~$n$ Boolean variables and~$m$
clauses in an instance. Each clause is chosen independently of the
others. Each variable in each clause is negated with a probability
of ${1}/{2}$, independently of the other variables. The model
differs from the regular model in several aspects: There is a community
structure on the set of variables, and we also do not necessarily
assume all clauses to be of the same length. Specifically, the set
of variables~$\left\{ v_{1},\dots,v_{n}\right\} $ is partitioned into
$B$ disjoint (sets of variables referred to as) communities $\mathcal{C}_{1},\mathcal{C}_{2},...,\mathcal{C}_{B}$.
For simplicity, we assume all communities to be of the same size $h$,
so that~$n=B\cdot h$. Without loss of generality, we will assume that $\mathcal{C}_i=\lbrace{v_j:(i-1)h+1\le j\le ih\rbrace}$, $1\le i\le B$. As~$n$ grows, so do usually both~$B$ and~$h$
(although at times one of them may remain fixed), and we will write~$B(n)$
and~$h(n)$ when we want to relate to their dependence on $n$. For
an~$\ell$-tuple~${\K}=(k_{1},\ldots,k_{\ell})$ with non-increasing, positive
integer entries, denote by~$\Omega_{B}\left(n,\:{\bf{K}}\right)$
the set of all clauses of length $k_1+\dots+k_{\ell}$, formed of~$k_{1}$ variables from some community~$\mathcal{C}_{i_{1}}$,~$k_{2}$
from another community~$\mathcal{C}_{i_{2}}$, $\ldots,$~$k_{\ell}$
from some $\ell$-th community~$\mathcal{C}_{i_{\ell}}$, where the indices $i_{j}$ are mutually distinct. We will refer to such a clause as a {\emph clause of type} ${\bf{K}}$. We will always implicitly assume
that $k_{i}\le h$ for each~$i$, so that we can indeed choose the required number of variables from the various communities. Similarly, we implicitly assume that $\ell\le B$. 
Let~$P_{{\bf{K}}}$ be the uniform
measure on~$\Omega_{B}\left(n,\:\K\right)$. 

\begin{example}\label{example1}
\begin{description}
\item{(a)} Let $n=1000$, $B=10$ and $h=100$.  The clauses 
$$( \overline{v}_{237}\lor \overline{v}_{250}\lor  v_{911}\lor  \overline{v}_{917}\lor v_{939}),$$ and
$$(v_{401}\lor \overline{v}_{423}\lor  v_{427}\lor \overline{v}_{450}\lor v_{500})$$ 
are of types $(3,2)$ and    $(5)$ (single-community clause), respectively.  (In general, the type of single-community clauses of length $k$ will be written as $(k)$.) The clauses above belong to the spaces $\Omega_{10}\left(1000,\:(3,2)\right)$ and $\Omega_{10}\left(1000,\:(5)\right)$, respectively. 
\item{(b)} The space  $\Omega_{10}\left(1000,\:(3,2)\right)$ consists of  $a=10\cdot 9\tbinom{100}{3}\tbinom{100}{2}\cdot 2^5$ clauses, and $\Omega_{10}\left(1000,\:(5)\right)$ consists of $b=10\tbinom{100}{5}\cdot 2^5$ clauses. 
\item{(c)} Under the measure $P_{(3,2)}$, each clause in $\Omega_{10}\left(1000,\:(3,2)\right)$ is chosen with probability $1/a$; under $P_{(5)}$, each clause in $\Omega_{10}\left(1000,\:(5)\right)$ is chosen with probability $1/b$.
\end{description}
\end{example}

The random instances we will be dealing with are of the following structure. There is some number $T\ge1$ of clause types ${\K}_{1},\ldots,{\K}_{T}$. Each ${\K}_{t}, 1\le t\le T$, is a vector ${\bf{K}}_{t}=(k_{1t},\ldots,k_{\ell t})$. These vectors are mutually distinct. Each clause in the instance is of one of these types. The probability of each clause to be of type ${\K}_{t}$ is $p_t$, where~$p_{t},1\le t\le T$, are arbitrary fixed real numbers, with~$\sum_{t=1}^{T}p_{t}=1$. More formally, we select independently $m$ clauses from the space~$\bigcup_{t=1}^{T}\Omega_{B}\left(n,\:{\bf{K}}_{t}\right)$ according to 
the measure~$P=\sum_{t=1}^{T}p_{t}\cdot P_{{\bf{K}}_{t}}$.
Using similar notations to \citep{Jordy 2015 modularity}, denote by~$F\left(n,m,B,P\right)$
the probability space of instances. Namely, the sample space of~$F\left(n,m,B,P\right)$ is the $m$-fold Cartesian product $\left(\bigcup_{t=1}^{T}\Omega_{B}\left(n,\:{\bf{K}}_{t}\right)\right)^{m}$ of the space corresponding to the selection of a single clause, endowed with the product measure~$P^{m}$. (For more on the notions of a product of measure spaces and of the product measure, see, for example, \citep[Sec. 2.5]{AdamsGuillemin}.)
For concreteness, in Algorithm~\ref{Algorithm} we present the exact mechanism for selecting an instance of~$F(n,m,B,P)$.
 
 \begin{algorithm}[ht]
\KwIn{$n$,  $m$, $B$,  ${\K}_{1},\ldots,{\K}_T$, $p_1,\ldots,p_T$.}
\KwOut{An instance $\I$}
$\I \leftarrow \emptyset$\;

 \For{$i\gets1$ \KwTo $m$}
 {$C \leftarrow \emptyset$\;
 Choose a clause type -- each ${\K}_t$ has probability $p_t$\;
 Suppose ${\K}_t=(k_1,\ldots,k_{\ell})$\;
 Select $\ell$ distinct integers $i_1,\ldots, i_\ell$ in the range $[1,B]$ (with the same probability     $\frac{1}{B(B-1)\cdots (B-\ell+1)}$ for each possible choice)\;
  \For{$j\gets1$ \KwTo $\ell$}
 {Choose $k_j$ distinct integers $a_1,\ldots,a_{k_j}$ in the range $[1,h]$ (with the same probability $\frac{1}{h(h-1)\cdots (h-k_j+1)}$ for each possible choice)\;
 \For{$d\gets1$ \KwTo $k_j$}
 {
 $x\leftarrow (i_j-1)\cdot h+a_d$\;
 Negate $v_x$ with probability $1/2$\;
 $C\leftarrow C\lor v_x$\;}}
$\I \leftarrow \I\land C$\;
   }{return $\I$}
    \caption{{\bf Choosing an instance in $F(n,m,B,P)$} \label{Algorithm}}

\end{algorithm}  
   
Note that, when employing Algorithm \ref{Algorithm}, we care about the order of choices, so that each clause may be obtained in several ways. This is easier to implement and has no bearing on the probability of obtaining each possible clause.
   
Thus, the regular model of random $k$-SAT is, with the notations above, $F\left(n,m,1,P_{(k)}\right)$. Instances in the model presented in \citep{Jordy 2015 modularity} include clauses of length $k$ of two types: ($i$) single-community clauses -- all $k$ variables belong
to the same community ($B\tbinom{h}{k}\cdot 2^k$ possible choices), and ($ii$) the $k$ variables belong to $k$ distinct
communities ($\tbinom{B}{k}\cdot (2h)^k$ possible choices). For some~$0<p<1$, each clause is of type ($i$) with probability $p$ and of type ($ii$) with probability~$1-p$. With the above notations, their probability space is\[
F\bigg(n,m,B,p\cdot P_{(k)}+(1-p)\cdot P_{\big(\underbrace{1,\ldots ,1}_{k}\big)}\bigg)
\]
for some $k$.


\begin{example}
With $n,B$ and $h$ as in Example \ref{example1}, and $k=3$, the instance 
$$(\overline{v}_{423}\lor v_{459}\lor v_{496})\land(v_{156}\lor\overline{v}_{437}\land v_{626}),$$
is an instance in 
\[
F\left(1000,2,10,0.2\cdot P_{(3)}+0.8\cdot P_{(1,1,1)}\right).
\]
The first clause is of type $(3)$ as all three variables ${v}_{423}, v_{459}, v_{496}$ belong to the same community $\mathcal{C}_5=\{v_i: 401\le i\le 500\}$, while the second clause is of type $(1,1,1)$, as the variables $v_{156}, {v}_{437}, v_{626}$ belong to three distinct communities: $\mathcal{C}_2$, $\mathcal{C}_5$ and $\mathcal{C}_7$, respectively. 
\end{example}

As our interest in this paper is in instances constructed as above, from this point on we will use the term ``community-structured" instead of the more general ``industrial".

\section{\label{sec:The-main-results}The Main Results}
As explained above, the clauses in an community-structured instance tend to include variables from the same community. 
In this paper, moreover, we usually deal
with the case where one (or more) of the clause types is a single-community type,
namely ${\bf{K}}_{t}=(k)$ for some~$1\le t\le T$ and~$k\le h$. In some results, we will
further restrict ourselves to the case~$T=1$, where the only clause type is a single-community type (equivalently, $P=P_{(k)}$ for some $k$).

In \citep{Jordy 2015 modularity}  it was observed empirically that,
when the modularity of the variable incidence graph  of the instance
increases, the threshold decreases. Now, the modularity in our case
is larger when more clauses consist of variables from the same community
and when the communities are small. Our first result is quite straightforward,
but it already hints that instances in the model suggested in Section
\ref{newcom sec:new model} tend to be no more satisfiable than random~$k$-SAT
instances. Note that the first part of the proposition is one of the
initial results for random SAT \citep{threshold franco} .
\begin{prop}
\label{prop3.2} Let $\mathcal{I}$ be a random instance in $F\left(n,\alpha n,B,P\right)$.
\begin{description}
\item{(a)} Suppose that for each ${\K}_{t}$, $1\le t\le T$, the
clause length is at most $k$. If $\alpha>2^{k}\ln2$, then $\mathcal{I}$
is UNSAT w.h.p.

\item{(b)} Let $T=1$ and $P=P_{(k)}$ for some $k\ge2$.
\begin{description}
\item{(i)} If $\alpha>r_{k}^{*}$, then $\mathcal{I}$ is UNSAT
w.h.p. 

\item{(ii)} If $h(n)=\Theta(n)$ and $\alpha<r_{k}$, then
$\mathcal{I}$ is SAT w.h.p.

\end{description}
\end{description}
\end{prop}

Our next result points out a significant difference between random
instances and community-structured ones. One might expect the threshold to be
different for community-structured instances, but it turns out that this difference
may be not just quantitative. The following result shows that, surprisingly,
under certain conditions the satisfiability threshold is~$0$. To
this end, we will consider~$m$ as some function of~$n$, not necessarily~$m=\alpha n$,
and write~$m(n)$ instead of~$m$. 

For real functions $f$ and $g$, we write $f=\Omega(g)$ if~$g=O(f)$,
and~$f=\omega(g)$ if~$g=o(f)$. We also write~$f=\textmd{polylog}(g)$
if~$f=O\left(\ln^{\theta}g\right)$ for some~$\theta$. 
\begin{thm}
\label{theorem1}Let $\mathcal{I}$ be a random instance in $F\left(n,m(n),B,P\right)$,
where ${\bf{K}}_{t}=(k)$ for some~$1\le~t\le~T$. 

\begin{description}
\item{(a)}
Let $h(n)=O\left(1\right)$.
\begin{description}
\item{(i)} If $T=1$ (so that $P=P_{(k)}$) and $m(n)=o(n^{1-1/2^{k}})$, then $\mathcal{I}$ is SAT w.h.p. 
\item{(ii)} If $m(n)=\Theta(n^{1-1/2^{k}})$, then $\mathcal{I}$ is SAT with probability bounded away  from $1$. If, moreover,  $T=1$, then $\mathcal{I}$ is SAT with probability bounded away   from both  $0$ and $1$.
\item{(iii)}  If $m(n)=\omega(n^{1-1/2^{k}})$, then $\mathcal{I}$
is UNSAT w.h.p.
\end{description}
\item{(b)} If $h(n)=o\left(\frac{\ln n}{\ln\ln n}\right)$ and $m(n)=\Omega\left(\frac{n}{\textnormal{\textmd{polylog}}(n)}\right)$,
then $\mathcal{I}$ is UNSAT w.h.p.
\item{(c)} If $h(n)=o\left(\ln n\right)$ and $m(n)=\Omega\left(n\cdot e^{-\beta\cdot\ln n/h(n)}\right)$
for some $\beta<1/r_{k}^{*}$, then $\mathcal{I}$ is UNSAT w.h.p.

\item{(d)} Let $h(n)=O\left(\ln n\right)$ and $T=1$. Then there exists
some~$\varepsilon_{0}>0$ such that, if $m(n)=\alpha n$ with~$\alpha> r_{k}^{*}-\varepsilon_{0}$, then $\mathcal{I}$ is UNSAT w.h.p.
\end{description}
\end{thm}

\newcommand\VRule[1][\arrayrulewidth]{\vrule width #1}
\setcellgapes{7pt}
\begin{centering}
\begin{table}[ht]
\makegapedcells
\centering
\begin{tabular}{!{\VRule[2pt]}c!{\VRule}c!{\VRule}c!{\VRule}!{\VRule}c!{\VRule[2pt]}}
\specialrule{2pt}{0pt}{0pt}
\multicolumn{3}{!{\VRule[2pt]}c!{\VRule}!{\VRule}}{\large{Parameters}} & \multicolumn{1}{c!{\VRule[2pt]}}{\large{Result}} \\[-.08ex]  \hhline{---}
\boldmath{$h(n)$} & \boldmath{$m(n)$} & \boldmath{$T$}& \\
\hline
& & & \\[-2.6ex] \hline
$O(1)$  & $o\left(n^{1-1/2^{k}}\right)$  & $1$ &  SAT w.h.p. \\\hline
$O(1)$  & $\Theta\left(n^{1-1/2^{k}}\right)$ & $1$  & $\in(\delta,1-\delta)$\\\hline
$O(1)$  & $\Theta\left(n^{1-1/2^{k}}\right)$ & $*$  & $\in(0,1-\delta)$\\\hline
$O(1)$ & $\omega\left(n^{1-1/2^{k}}\right)$ & $*$  & {UNSAT  w.h.p.} \\
\hline
$o\left(\frac{\ln n}{\ln\ln n}\right)$ & $\Omega\left(n/\textmd{polylog}(n)\right)$ & $*$ &  {UNSAT  w.h.p.} \\\hline
$o\left(\ln n\right)$  & $\Omega\big(n^{1-1/(r_{k}^{*}+\varepsilon)h(n)}\big)$ & $*$ & {UNSAT  w.h.p.}  \\\hline
$O\left(\ln n\right)$ & $>\left(r^{*}_{k}-\varepsilon_0 \right) n$ & $1$ &  
{UNSAT  w.h.p.}  \\ \specialrule{2pt}{0pt}{0pt}
\end{tabular}
\caption{Asymptotic satisfiability of a random instance with small communities in~$F\left(n,m(n),B,P\right)$. }
\label{table1}
\end{table}
\end{centering}

\begin{rem}
\label{remark 1}
\begin{description}
\item{(a)} The $\varepsilon_{0}$ in part (d) is effective. Namely,
as will follow from the proof, one can present such an~$\varepsilon_{0}$
explicitly (in terms of the implicit constant in
the equality $h(n)=O(\ln n)$).

\item{(b)} Still in case (d), one can deal with the general case of arbitrary
$T$ as long as the weight of $P_{(k)}$ in~$P$, namely the probability that a clause is of type $P_{(k)}$, is sufficiently large.
\end{description}
\end{rem}

In Theorem \ref{theorem1} there are four types of results for the asymptotic
satisfiability of a random community-structured instance with~$n$ variables,
$m(n)$ clauses, $B$ communities of size~$h(n)=n/B$, and probability measure $P$. Namely,
either the probability of satisfiability (i) tends to~$0$ as~$n\to\infty$,
or (ii) it tends to $1$, or (iii) it is bounded away from~$1$, or (iv) it is bounded away from both~$0$ and~$1$.
These results are summarized in Table \ref{table1}. In general, we assume that~${\bf{K}}_{t}=(k)$ for some~$1\le t\le T$ and $k\ge 1$. In the third column we place a `$1$' or a `$*$', depending on whether $T$ is required to be $1$ or is arbitrary, respectively.
The notation $\in(0,1-\delta)$  indicates a
probability bounded away from~$1$, and the notation $\in(\delta,1-\delta)$ indicates a
probability bounded away both from~$0$ and~$1$.


The proof of Theorem \ref{theorem1} will use the following lemma.
\begin{lem}
\label{lemma3.4}Consider the spaces $F\left(n,m(n),B,P\right)$ and $F\left(n,m'(n),B,P\right)$, where $m'(n)=\omega\left(m(n)\right)$.
If a random instance in $F\left(n,m(n),B,P\right)$ is UNSAT with
probability bounded away from $0$, then a random instance
in~$F\left(n,m'(n),B,P\right)$ is UNSAT w.h.p. 
\end{lem}
In the proof of Theorem \ref{theorem1} (and that of Theorem \ref{theorem2}), we use some results regarding the classical ``balls and
bins'' problem. In this problem, there
are $M$ balls and~$B$ bins. Each ball is placed uniformly randomly
in one of the bins, independently of the other balls. One quantity
of interest is the $\textit{maximum load}$, which is the maximum
number of balls in any bin. There are several papers studying the
size of the maximum load, as well as generalizations of this problem.
It seems that \citep{balls and bins Raab n Steger} contains all previous
results. Our next result seem  not to be covered by previous results
regarding the balls and bins problem. It  will be employed in the proof
of Theorem \ref{theorem1}, and is of independent interest. 

Given a sequence $(X_n)_{n=1}^{\infty}$ of random variables and a probability law $\L$, we let $X_n\xrightarrow[n\to\infty]{\D}\L$ denote the fact that $X_n$ converges to $\L$ in distribution as $n\to\infty$. Denote by $\Po(\lambda)$ the Poisson distribution with parameter $\lambda$.

\begin{thm}
\label{thm3.5} Consider the balls and bins problem with~$B$ bins
and $M=M(B)$ balls, where $B\rightarrow\infty$. Let~$s\ge 2$ be an arbitrarily fixed integer.
\begin{description}
\item{(a)} If $M(B)=o(B^{1-1/s})$, then the maximum load is at most $s-1$ w.h.p.

\item{(b)} If $M(B)=\omega(B^{1-1/s})$, then the maximum load is at least $s$ w.h.p.

\item{(c)} If $M(B)=\Theta(B^{1-1/s})$, then the maximum load is either $s-1$ or $s$ w.h.p. Moreover, suppose
\begin{equation}
M(B)=C\cdot B^{1-1/s}\left(1+o(1)\right)\label{eqn1},
\end{equation}
and let $X_B$ be the number of bins that contain exactly $s$ balls. Then $X_B\xrightarrow[B\to\infty]{\D}\Po\mathit{(C^s/s!)}$.
\end{description}

\end{thm}
Theorem \ref{thm3.5} will be proven in Appendix \ref{appendixa}.

As noted earlier, random $2$-SAT is much better understood than random
$k$-SAT for general $k$. This enables us to obtain a stronger result
than Theorem \ref{theorem1} in the case $P=P_{(2)}$. 


\begin{thm}
\label{theorem2}Let~$\mathcal{I}$ be a random instance in $F\left(n,\alpha n,B,P_{(2)}\right)$.
\begin{description}
\item{(a)} There exists an $0<\varepsilon_{0}<1$ such that, if $h(n)=o\left(\sqrt{n}\right)$
and~ $\alpha>1-\varepsilon_{0}$, then $\mathcal{I}$ is UNSAT w.h.p. 

\item{(b)} For $h(n)=\Theta(\sqrt{n})$: 
\begin{description}
\item{(i)} If $1-\varepsilon_{0}<\alpha<1$, where $\varepsilon_{0}$
is as in $(a)$, then $\mathcal{I}$ is SAT with probability bounded
away from both $0$ and $1$.

\item{(ii)} If $\alpha=1$ then $\mathcal{I}$ is UNSAT w.h.p.

\end{description}

\item{(c)} For $h(n)=\omega(\sqrt{n})$ with $h(n)=o(n)$:
\begin{description}
\item{(i)} If $\alpha<1$ then $\mathcal{I}$ is SAT w.h.p. 

\item{(ii)} If $\alpha=1$ then $\mathcal{I}$ is UNSAT w.h.p.

\end{description}

\item{(d)} For $h(n)=\Theta(n)$ :
\begin{description}
\item{(i)} If $\alpha<1$ then $\mathcal{I}$ is SAT w.h.p. 

\item{(ii)} If $\alpha=1$ then $\mathcal{I}$ is SAT with
probability bounded away from both $0$ and $1$.
\end{description}
\end{description}
\end{thm}

\begin{rem} As in Remark \ref{remark 1}.(b), one can deal with the more
general case of arbitrary $T$, as long as one of the clause types ${\bf{K}}_{t}$ is of the form  $(2)$ and is of sufficiently large weight.\end{rem}
Similarly to Table \ref{table1}, we summarize the results of Theorem \ref{theorem2} in Table \ref{table2}. Here, we always assume $k=2$,~$m(n)=\alpha n$ and~$T=1$. 
The notations are as in Table \ref{table1}.
\begin{centering}
\setcellgapes{7pt}
\begin{table}[ht]
\makegapedcells
\centering
\begin{tabular}{!{\VRule[2pt]}c!{\VRule}c!{\VRule}c!{\VRule[2pt]}}
\specialrule{2pt}{0pt}{0pt}
\multicolumn{1}{!{\VRule[2pt]}c!{\VRule}}{\diaghead(-4,1){aaaaaaaaaaaaaaaaa}{\boldmath{$h(n)$}}{\boldmath{$\alpha$}}}&
$\in(1-\varepsilon_{0},1)$ & $ =1$ \\ \hline
& & \\[-2.6ex] \hline
$=o\left(\sqrt{n}\right)$ & $\textnormal{UNSAT w.h.p}$  & $ \textnormal{UNSAT w.h.p.}$\\\hline
$=\Theta\left(\sqrt{n}\right)$ & $\in(\delta ,1-\delta )$  & $ \textnormal{UNSAT w.h.p.}$\\\hline
$\in \omega\left(\sqrt{n}\right) \cap o\left(n\right)$ & $\textnormal{SAT w.h.p.}$  & $ \textnormal{UNSAT w.h.p}$\\\hline
$=\Theta\left(n\right)$ & $\textnormal{SAT w.h.p}$  & $ \in(\delta ,1-\delta )$\\
\specialrule{2pt}{0pt}{0pt}
\end{tabular}
\caption{Asymptotic satisfiability of a random instance in~$F\left(n,\alpha n,B,P_{(2)}\right)$. }
\label{table2}
\end{table} 
\end{centering}

As we have seen, when clauses tend to be formed of variables in the same community, the instance tends to become unsatisfiable. One may wonder what happens in the opposite case, namely when variables tend to belong to distinct communities. Intuitively, this constraint should usually make little difference, as anyway few clauses may be expected to contain variables from the same community. However, if there are very few communities, this constraint is more significant. Specifically, consider the extreme case of $B=2, P=P_{(1,1)}$. In this case, we disallow about half of the possible clauses of the classical model $B=1, P=P_{(2)}$. Does it affect the satisfiability threshold? Namely, if when moving from the classical case to a case with most clauses from the same community, we tend to make the instance unsatisfiable, will the constraint of having in each clause variables from distinct communities tend to make it ``more satisfiable"? The following theorem shows that it makes a very small difference if at all. 

\begin{thm}\label{thm3.9}
Let $0\leq p\leq 1$ and $B\geq 2$ arbitrary and fixed. The satisfiability threshold in the model 
$$F\left(n,m,B, pP_{(1,1)}+(1-p)P_{(2)}\right)$$
is $1$.
\end{thm}

One may still ask whether the regular random model $F(n,m,1,P_{(2)})$ and the model $F(n,m,2,P_{(1,1)})$ display some difference in behaviour near the threshold, namely for $m=n\cdot(1+o(1))$. More precisely, recall that, by  \citep{scaling window}, for $m$
 in some range of  size $\Theta(n^{2/3})$ around $n$, the satisfiability probability for the random model is bounded away from both $0$ and $1$. (See (\ref{eq:window 1}) below for a more accurate formulation.) Do the two models behave in the same way for $m=n+\theta n^{2/3}$ for fixed $\theta$?
 
 We studied this question by a large simulation. We detail the experiment in Section $5$. The results seem to indicate strongly that the two models behave in the same way also in the window~$m=n\pm \Theta(n^{2/3})$.
 
\section{Proofs}

\begin{Proof}{\ of Proposition \ref{prop3.2}} 
\begin{description}
\item[(a)] We follow the proof
in the random model \citep{threshold franco}. Fix a truth assignment
and consider~$\mathcal{I}$. Each clause has at most $k$ literals.
The variables are negated with probability $1/2$ independently of
each other, and hence each clause is satisfied with probability of
at most $1-2^{-k}$, independently of the other clauses. The expected number
of satisfying truth assignments is therefore at most $2^{n}\cdot\left(1-2^{-k}\right)^{\alpha n}$.
As $\alpha>2^{k}\ln2$, we have 
\[
2^{n}\cdot\left(1-2^{-k}\right)^{\alpha n}\xrightarrow[n\to\infty]{}0.
\]
(Note that we have not used in this part the specific mechanism by which clauses are selected. The variables in each clause may be selected arbitrarily. As long as all clauses are of length at most $k$, and the sign of each variable in each clause is selected uniformly randomly, and independently of all other variables in this clause and all the others, the conclusion holds.)  Thus, $\mathcal{I}$ is UNSAT w.h.p.
 
 \item[(b)] A random instance~$\mathcal{I}$ in $F\left(n,\alpha n,B,P_{(k)}\right)$
decomposes into $B$ sub-instances~$\mathcal{I}_{i}$,~$1\le i\le B$,
where each~$\mathcal{I}_{i}$ is formed of those clauses consisting
of variables solely from $\mathcal{C}_{i}$. Obviously,~$\mathcal{I}$~is SAT if and only if all~$\mathcal{I}_{i}$-s are such. For~$1\le i\le B$,
let~$U_{i}=1$ if~$\mathcal{I}_{i}$ is satisfiable, and~$U_{i}=0$
otherwise. The variable~$U=\prod_{i=1}^{B}U_{i}$ indicates whether~$\mathcal{I}$
is satisfiable. Let~$W_{i}$ denote the number of clauses in~$\mathcal{I}_{i}$. Since each of the $\alpha n$ clauses consists of variables from~$\mathcal{C}_{i}$  with probability $1/B$, independently of all other clauses, $W_{i}$~is binomially distributed with parameters $\alpha n$ and $1/B$:
\[
W_{i}\sim B\left(\alpha n,1/B\right),\qquad1\le i\le B.
\]

\begin{description}

\item[(i)] Suppose first that $h(n)=\omega(1)$. Let~$\alpha_{i}$ denote
the density of the sub-instance $\mathcal{I}_{i}$, $1\le i\le B$.
There exists an $i$ with~$\alpha_{i}\ge\alpha$,~and therefore
$\alpha_{i}>r_{k}^{*}$. It follows that $\mathcal{I}_{i}$ is UNSAT
w.h.p., and hence so is $\mathcal{I}$. 

The case $h(n)=O(1)$ follows in particular from Theorem \ref{theorem1}.(a).(iii) (to be proved below).

\item[(ii)] In this case $B(n)=\Theta(1)$. Without loss of generality assume
$B(n)=B$ is fixed. For $\gamma>0$, let 
\begin{align}
W_{<\gamma}=\bigcap_{i=1}^{B}\left\{ W_{i}<\gamma\cdot h(n)\right\} .\label{proposition 3.2 part a ii label 0}
\end{align}
Let $\alpha'$ be an arbitrary fixed number, strictly between $\alpha$
and $r_{k}$. Denoting by $\overline{A}$ the complement of an event $A$, we have:
\begin{equation}
\begin{aligned}
P\left(U=1\right)= & \:P\left(W_{<\alpha'}\right)P\left(U=1\left|W_{<\alpha'}\right.\right)\\
 & \:+P\left(\overline{W}_{<\alpha'}\right)P\left(U=1\left|\overline{W}_{<\alpha'}\right.\right)\\
 \ge & \:P\left(W_{<\alpha'}\right)P\left(U=1\left|W_{<\alpha'}\right.\right).
\end{aligned}
\label{proposition 3.2 part a ii label 1}
\end{equation}
By the weak law of large numbers for the binomial random variables $W_i$, 
\[
\dfrac{W_{i}}{n/B}\xrightarrow[n\to\infty]{P}\alpha,\qquad1\le i\le B,
\]
and therefore 
\[
\begin{aligned}P\left(\overline{W}_{<\alpha'}\right) & =P\left(\bigcup_{i=1}^{B}\left\{ W_{i}\ge\alpha'\cdot n/B\right\} \right)\\
 & \le\sum_{i=1}^{B}P\left(W_{i}\ge\alpha'\cdot n/B\right)\\
 & =B\cdot P\left(W_{1}\ge\alpha'\cdot n/B\right)\xrightarrow[n\to\infty]{}0.
\end{aligned}
\]
 Hence
\begin{align*}
P\left(W_{<\alpha'}\right)=1-P\left(\overline{W}_{<\alpha'}\right) & \xrightarrow[n\to\infty]{}1.
\end{align*}
 Now consider the second factor on the right-hand side of (\ref{proposition 3.2 part a ii label 1}).
Clearly, the more clauses any $\I_i$ contains, the less likely it is to be satisfiable, and therefore
\begin{equation}
\begin{aligned}
P\left(U=1\left|W_{<\alpha'}\right.\right) & \ge\prod_{i=1}^{B}P\left(U_{i}=1\left|W_{i}=\alpha'\cdot n/B\right.\right).
\end{aligned}
\label{proposition 3.2 part a ii label 4}
\end{equation}

As $\alpha'<r_{k}$, each of the sub-instances ${\mathcal I}_i$ is SAT w.h.p., so that each of the factors in the product on the right-hand side of (\ref{proposition 3.2 part a ii label 4}) converges to~1 as $n\to\infty$. Since, by our assumption, $B$ is fixed, so does the whole product. Hence~$\mathcal{I}$ is SAT w.h.p.

\end{description}
\end{description}
\qed
\end{Proof}
\bigskip{}

As mentioned in Section \ref{sec:The-main-results}, the proofs of
Theorem \ref{theorem1} and Theorem \ref{theorem2} make use
of some results concerning the balls and bins problem. Let $L$ be
the maximum load for $M$ balls and $B$ bins. By \citep{balls and bins Raab n Steger},
for any~$\delta>0$, 
\begin{equation}
L\ge\begin{cases}
\dfrac{\ln B}{\ln\frac{B\ln B}{M}}, & \:\textnormal{if }\frac{B}{\textmd{polylog}(B)}<M=o(B\ln B),\\
\left(d_{c}-\delta\right)\ln B, & \: \textnormal{if }M=c\cdot B\ln B,
\end{cases}\label{balls and bins 1}
\end{equation}
w.h.p.\,for an appropriate constant $d_{c}>c$.

\begin{rem}
 The constant $d_{c}$, in the second part of (\ref{balls and bins 1}),
is the unique solution of the equation 
\[
1+x(\ln c-\ln x+1)-c=0
\]
in $(c,\infty)$ (see \citep[Lemma 3]{balls and bins Raab n Steger}).
A routine calculation shows that $d_{c}=c+c\cdot u(1/c)$, where the
function~$u$ is the unique non-negative function defined implicitly
by the equation 
\[
-u(w)+(1+u(w))\ln(1+u(w))=w,\qquad(w\ge0).
\]
The function $u(w)$ has been studied in \citep[pp.\ 101\textendash 102]{allocation},
and in particular expressed as a power series in $\sqrt{w}$ near
$0$.

The fact that $d_{c}>c$ is the reason that the threshold 
in Theorem \ref{theorem1}.(d) is strictly less than  $r_{k}^{*}$. One can
easily bound~$d_{c}-c$ from below. In fact, write~$d_{c}=c+\varepsilon$.
Then
\begin{align*}
1 & =-(c+\varepsilon)(\ln c-\ln(c+\varepsilon)+1)+c\\
 & =-(c+\varepsilon)\ln c+(c+\varepsilon)\ln(c+\varepsilon)-\varepsilon\\
 & <(c+\varepsilon)\cdot \varepsilon/c-\varepsilon=\varepsilon^{2}/c,
\end{align*}
and hence $d_{c}>c+\sqrt{c}$.
\label{remark41}
\end{rem}

\begin{Proof}{\ of Theorem \ref{theorem1}} 
We follow the notations used in the proof of Proposition \ref{prop3.2}. Recall that~$\mathcal{I}_{i}$ is the sub-instance formed of those clauses
in $\mathcal{I}$ consisting of variables solely from $\mathcal{C}_{i}$,
and~$W_{i}$ is the number of clauses in~$\mathcal{I}_{i}$,~$1\le i\le B$.
Denote $W_{\max}=\max\left\{ W_{1},\ldots,W_{B}\right\} .$

Note that, while we have not assumed that $T=1$  in parts (a).(iii), (b), and (c) of the theorem, we may make this assumption without loss of generality in these parts as well.
In fact, suppose that any of these three parts  has been proven for the case $T=1$, and consider the general case. If the probability of  $(k)$ in~$P$
is $p$, then w.h.p.\ there will be at least $p\cdot m(n)/2$
clauses  of type $(k)$. To see this, denote by $\mathcal{I}^{(k)}$ the sub-instance of $\mathcal{I}$ obtained by taking the clauses of type $(k)$ and by $m'(n)$ the number of clauses in~$\mathcal{I}^{(k)}$. Clearly, $m'(n)$ is distributed binomially with parameters $m(n)$ and~$p$. Employing Chernoff's bound we obtain a lower bound of $p\cdot m(n)/2$ on~$m'(n)$ w.h.p.  It follows that $m'(n)$ has the same lower bound assumed on $m(n)$ (namely, it is $\omega(n^{1-1/2^{k}})$ in part (a).(iii), it is $\Omega\left(\frac{n}{\textnormal{\textmd{polylog}}(n)}\right)$ in part (b), and it is $\Omega\left(n\cdot e^{-\beta\cdot\ln n/h(n)}\right)$ in part (c)). As we have assumed the correctness of these parts for $T=1$, the instance $\mathcal{I}^{(k)}$ is  UNSAT w.h.p., and hence certainly so is the original instance, which contains it. Thus, we may indeed assume in all parts that $T=1$.

Since $T=1$, each clause has all its literals from the same community. Hence, the selection of a clause corresponds to the selection of a community.
Consider clauses as balls, and communities as bins. 
The process of selecting the clauses, as far as the community to which the variables in each
clause belong, is analogous to that of placing $m(n)$ balls in~$B$
bins uniformly at random. The idea of the proof in parts (b)-(d) will
be to prove that w.h.p.\ we have $W_{\max}/h(n)>r_{k}^{*}$. This
will imply that there is at least one sub-instance~$\mathcal{I}_{i}$
with density larger than~$r_{k}^{*}$. Thus, already~$\mathcal{I}_{i}$
is UNSAT w.h.p., and consequently so is~$\mathcal{I}$.

\begin{description}
\item[(a)] Without loss of generality, assume that~$h(n)=h>0$
is fixed. 
\begin{description}
\item[(i)] By Theorem \ref{thm3.5}.(a), there is no sub-instance with more
than~$2^{k}-1$ clauses w.h.p. Since instances with less than $2^{k}$
clauses are certainly satisfiable, all $\mathcal{I}_{i}$-s
are SAT, and hence so is $\mathcal{I}$.

\item[(ii)] Here, we may assume that  $m(n)=\theta\cdot n^{1-1/2^k}$ for some constant $\theta>0$. By Theorem \ref{thm3.5}.(c), the probability that there is
an~$\mathcal{I}_{i}$,~$1\le i\le B$, with at least~$2^{k}$ clauses,
is bounded away from~$0$. Assume, say, that~$W_{1}\ge2^{k}$. Then,
with probability at least$$\left(1/\binom{h}{k}\right)^{2^{k}}\cdot(2^{k})!/2^{k2^{k}},$$
all ~$2^{k}$ distinct clauses consisting of the variables~$v_{1},\ldots,v_{k}$
have been drawn. As the instance is UNSAT if it contains all these~$2^{k}$
clauses, the probability for our instance to be UNSAT is bounded away
from~$0$. 
Now, assume that $P=P_{(k)}$, for some $k>0$. Now, by Theorem \ref{thm3.5}.(c) there is no sub-instance with more
than~$2^{k}-1$ clauses  with probability bounded away from $0$. Thus, similarly to part (i),  $\mathcal{I}$ is SAT with probability bounded away from $0$.

\item[(iii)] Follows from the previous part and Lemma \ref{lemma3.4}.

\end{description}

\item[(b)] In view of part (a).(iii), we may assume $h(n)\to\infty$. We may
also assume that ~$m(n)=n/\ln^{\theta}n$ for some~$\theta\ge1$.
Clearly, $m(n)\le B$. On the other hand, 
\[
m(n)=\dfrac{n}{\ln^{\theta}n}\ge\dfrac{B}{(2\ln B)^{\theta}}=\dfrac{B}{\textmd{polylog}(B)}.
\]
Thus, by (\ref{balls and bins 1}), w.h.p., the maximum load is at
least 
\begin{equation}
\begin{aligned}\dfrac{\ln B}{\ln\dfrac{B\ln B}{m(n)}} & \ge\dfrac{\frac{1}{2}\cdot\ln n}{\ln\left(\dfrac{n\cdot\ln n}{n/\ln^{\theta}n}\right)}\ge\dfrac{\ln n}{2(\theta+1)\ln\ln n}.\end{aligned}
\label{theorem 3.2(b) label 2}
\end{equation}
Now, there are $h(n)=o(\ln n/\ln\ln n)$ variables in each community.
By (\ref{theorem 3.2(b) label 2}), w.h.p., the density of the sub-instance~$\mathcal{I}_{i}$
with the maximal number of clauses is at least
\[
\begin{aligned}\dfrac{W_{\max}}{h(n)}\ge\dfrac{\frac{1}{2(\theta+1)}\cdot\ln n/\ln\ln n}{o\left(\ln n/\ln\ln n\right)} & \xrightarrow[n\to\infty]{}\infty.\end{aligned}
\]
Hence, this $\mathcal{I}_{i}$ is UNSAT w.h.p., and therefore so is
$\mathcal{I}$.

\item[(c)] By (\ref{balls and bins 1}), w.h.p., the number of clauses in
the sub-instance~$\mathcal{I}_{i}$ with the maximal number of clauses
is at least
\[
\begin{aligned}W_{\max} & \ge\dfrac{\ln B}{\ln\dfrac{B\ln B}{m(n)}}=\dfrac{\ln n(1-o(1))}{\ln\dfrac{B\ln B}{m(n)}}.\end{aligned}
\]
 For a large enough $n$
\[
\begin{aligned}
\ln\dfrac{B\ln B}{m(n)} & \le\ln\left(\dfrac{\dfrac{n}{h(n)}\cdot\ln n}{n\cdot e^{-\beta\ln n/h(n)}}\right)\\
 & =\ln\dfrac{\ln n}{h(n)}+\beta\cdot\dfrac{\ln n}{h(n)}.
\end{aligned}
\]
As $\beta<1/r_{k}^{*}$, for large enough $x$ we have $\ln x+\beta x<x/r_{k}^{*}.$
Hence, for large enough $n$ we have 
\[
\ln\dfrac{B\ln B}{m(n)}\le\dfrac{1}{r_{k}^{*}}\cdot\dfrac{\ln n}{h(n)}.
\]
This implies that the density of the sub-instance~$\mathcal{I}_{i}$
with the maximal number of clauses is at least
\[
\begin{aligned}\dfrac{W_{\max}}{h(n)}\ge\dfrac{1}{\ln\dfrac{B\ln B}{m(n)}}\cdot\dfrac{\ln n(1-o(n))}{h(n)} & >r_{k}^{*},\end{aligned}
\]
and thus UNSAT w.h.p. Consequently, so is $\mathcal{I}$.

\item[(d)] In view of the previous part, we may assume that $h(n)=\theta\ln n$
for some $\theta>0$. Choose $c_{0}$ such that $$d_{c_{0}}=\theta r_{k}^{*},$$  where $d_{c}$ is as in (\ref{balls and bins 1}). 
 Let $\alpha>c_{0}/\theta$, and put $c=\alpha \theta$. Thus, $c>c_{0}$ and~$d_{c}>d_{c_{0}}$. Let 
$\delta< d_{c}-\theta r_{k}^{*}$ .
We have 
\[
\begin{aligned}
m(n)=n\cdot\alpha=\dfrac{nc}{\theta} & =(1+o(1))\cdot\dfrac{nc\ln B}{\theta\ln n}\\
 & =(c+o(1))\cdot B\ln B.
\end{aligned}
\]
By (\ref{balls and bins 1}), the size of the largest sub-instance is 
$W_{\max}\ge(d_{c}-\delta)\ln B$ w.h.p.  Hence, w.h.p.\ the density of this sub-instance is
\[
\begin{aligned}\dfrac{W_{\max}}{h(n)} & \ge\dfrac{\left(d_{c}-\delta\right)\ln B}{h(n)}  =\dfrac{\left(d_{c}-\delta\right)\cdot(1-o(1))\ln n}{\theta\ln n}\\
 & =\dfrac{d_{c}-\delta}{\theta}-o(1)=r^{*}_{k}+\dfrac{d_{c}-\delta-\theta r^{*}_{k}}{\theta}-o(1).
 \end{aligned}
\]

Letting $\varepsilon_{0} =r_{k}^{*}-c_{0}/\theta$, we get our claim.
\qed
\end{description}
 \end{Proof}

\vspace{10pt}
\begin{Proof}{\ of Lemma \ref{lemma3.4}} Denote the random instance
in~$F\left(n,m'(n),B,P\right)$ by $\mathcal{I}'$. Denote the instance obtained
from the first~$m(n)$ clauses of~$\mathcal{I}'$ by $\mathcal{I}'_{1}$,
the instance obtained from the next~$m(n)$ clauses of~$\mathcal{I}'$
by $\mathcal{I}'_{2}$, $\ldots$ , the instance obtained from the~last
$m(n)$ clauses of~$\mathcal{I}'$ by~$\mathcal{I}'_{b(n)}$
(with $b(n)=m'(n)/m(n)$). According to our assumption, there exists
an $\varepsilon>0$ such that 
\[
P\left(\mathcal{I}'_{i}\textmd{ is SAT}\right)\le1-\varepsilon,\qquad i=1,2,\ldots,b(n).
\]
Now, the events $\left\{ \mathcal{I}'_{i}\textmd{ is SAT}\right\} ,1\le i\le b(n)$,
are independent, and we clearly have 
\[
\left\{ \mathcal{I}'\textmd{ is SAT}\right\} \subseteq\bigcap_{i=1}^{b(n)}\left\{ \mathcal{I}'_{i}\textmd{ is SAT}\right\} .
\]
 Since $b(n)\to\infty$:
\[
P\left(\mathcal{I}'\textmd{ is SAT}\right)\le P\left(\mathcal{I}'_{i}\textmd{ is SAT},1\le i\le b(n)\right)\le\left(1-\varepsilon\right)^{b(n)}\xrightarrow[n\to\infty]{}0.
\]

\qed

\end{Proof}  \bigskip


In the proof of Theorem \ref{theorem2} we will use the following
result from \citep{scaling window}. There exist some $0<\varepsilon_{0}<1$
and~$\lambda_{0}>0$ such that the satisfiability probability of a
random~$2$-SAT instance $\mathcal{I}$ with~$m=n\cdot(1+\varepsilon)$
clauses is 
\begin{equation}
P\left(\mathcal{I}\textmd{ is SAT}\right)=\begin{cases}
1-\Theta\left(\dfrac{1}{n\left|\varepsilon\right|^{3}}\right), & \:-\varepsilon_{0}\le\varepsilon\le -\lambda_{0}n^{-1/3},\\
\\
\Theta\left(1\right), & \:-\lambda_{0}n^{-1/3}<\varepsilon<\lambda_{0}n^{-1/3},\\
\\
\exp\left(-\Theta\left(n\varepsilon^{3}\right)\right), & \:\lambda_{0}n^{-1/3}\le\varepsilon\le\varepsilon_{0}.
\end{cases}\label{eq:window 1}
\end{equation}
Actually, in the sequel, we will encounter only the first two cases.
Note that in the case~$m=n\cdot(1-\varepsilon)$ with~$\lambda_{0}n^{-1/3}\le\varepsilon\le\varepsilon_{0}$,
we have

\begin{align}
1-\dfrac{\theta_{1}}{n\cdot\varepsilon^{3}}\le P\left(\mathcal{I}\textmd{ is SAT}\right) & \le1-\dfrac{\theta_{2}}{n\cdot\varepsilon^{3}}\label{eq:window 3}
\end{align}
for some constants $\theta_{1},\theta_{2}>0$.

We will also use an additional result regarding the balls and bins
problem. Let $L$ be the maximum load for~$M$ balls and~$B$ bins.
By \citep{balls and bins Raab n Steger}, w.h.p.
\begin{equation}
L\le\dfrac{M}{B}+\sqrt{\dfrac{2M\ln B}{B}},\qquad M=\omega(B\ln^{3}B).\label{balls and bins 2}
\end{equation}

Given a sequence $\left(X_{i}\right)_{i=1}^{\infty}$ of random variables
and a probability law $\mathcal{L},$ we write~$X_{i}\xrightarrow[i\to\infty]{\mathcal{D}}\mathcal{L}$
if the sequence converges to $\mathcal{L}$ in distribution.

\vspace{10pt}
\begin{Proof}{\ of Theorem \ref{theorem2}} We follow the notations
used in the proof of Theorem \ref{theorem1}. Also, for~$\gamma>0$,
let
\begin{align*}
I_{<\gamma}=\Bigl\{\left(m_{1},\ldots,m_{B}\right): & m_{1}+\ldots+m_{B}=\alpha n, m_{i}<\gamma\cdot h(n)\forall1\le i\le B\Bigr\},
\end{align*}
and let $I_{>\gamma}$ and $I_{\ge\gamma}$ be analogously understood.
More generally, for $0\le p\le1$, let~$I_{<\gamma,p}$ denote the
set of~$B$-tuples~$\left(m_{1},\ldots,m_{B}\right)$ with at least~$pB$
entries $m_{i}$,~$1\le i\le B$, for which~$m_{i}<\gamma\cdot h(n)$.
(Thus,~$I_{<\gamma}=I_{<\gamma,1}.)$ 

Note that the set $W_{<\gamma}$, defined in (\ref{proposition 3.2 part a ii label 0}),
may now be written in the form 
\[
W_{<\gamma}=\bigcup_{\left(m_{1},\ldots,m_{B}\right)\in I_{<\gamma}}\bigcap_{i=1}^{B}\left\{ W_{i}=m_{i}\right\} .
\]
 We use similar notations, for example $W_{>\gamma}$,~$W_{\ge\gamma,p}$
and~$W_{<\gamma,p}$, analogously.

\begin{description}
\item{(a)} Let $\delta,p$ be sufficiently small positive numbers, to be
determined later. Let $\varepsilon_{0}$ be as in (\ref{eq:window 1}).
We have
\begin{equation}
\begin{aligned}P\left(U=1\right)=
 & \:P\left(W_{<1+\delta}\cap W_{>1-\varepsilon_{0},p}\right)\cdot P\left(U=1\left|W_{<1+\delta}\cap W_{>1-\varepsilon_{0},p}\right.\right)\\
 & +P\left(W_{<1+\delta}\cap\overline{W}_{>1-\varepsilon_{0},p}\right)\cdot P\left(U=1\left|W_{<1+\delta}\cap\overline{W}_{>1-\varepsilon_{0},p}\right.\right)\\
 & +P\left(\overline{W}_{<1+\delta}\right)P\left(U=1\left|\overline{W}_{<1+\delta}\right.\right)\\
\le & \:P\left(U=1\left|W_{<1+\delta}\cap W_{>1-\varepsilon_{0},p}\right.\right)+P\left(W_{<1+\delta}\cap\overline{W}_{>1-\varepsilon_{0},p}\right)\\
  & +P\left(U=1\left|\overline{W}_{<1+\delta}\right.\right).
\end{aligned}
\label{com24 item b part 0}
\end{equation}

Consider the first term on the right-hand side of (\ref{com24 item b part 0}).
The event~$W_{<1+\delta}\cap W_{>1-\varepsilon_{0},p}$ implies that~$W_{j}=m_{j}$,~$1\le j\le B$,
for some~$\left(m_{1},\ldots,m_{B}\right)\in I_{<1+\delta}\cap I_{>1-\varepsilon_{0},p}$.
We note that, conditioned on the event~$\bigcap_{i=1}^{B}\left\{ W_{i}=m_{i}\right\} $,
the events~$\left\{ U_{i}=1\right\} $,~$1\le i\le B$, are independent.
Also, for each~$1\le i\le B$ with~$m_{i}>\left(1-\varepsilon_{0}\right)h(n)$
we have 
\begin{align*}
P\bigl(U_{i}=1\left|W_{i}=m_{i}\right.\bigr) & \le P\bigl(U_{i}=1\left|W_{i}=\left(1-\varepsilon_{0}\right)h(n)\right.\bigr).
\end{align*}
Thus 
\begin{equation}
\begin{aligned} P\left(U=1\left|W_{<1+\delta}\cap W_{>1-\varepsilon_{0},p}\right.\right) & \le \prod_{i:W_{i}>(1-\varepsilon_{0})h(n)}P\bigl(U_{i}=1\left|W_{i}=\left(1-\varepsilon_{0}\right)h(n)\right.\bigr)\\
 & \le P\bigl(U_{1}=1\left|W_{1}=\left(1-\varepsilon_{0}\right)h(n)\right.\bigr)^{pB}.
\end{aligned}
\label{com24 item b part 1}
\end{equation}
In view of Theorem \ref{theorem1}.(a).(iii), we may assume that~$h(n)\to\infty$.
By (\ref{eq:window 3}), for some~$\theta>0$
\begin{equation}
\begin{aligned}
P\left(U=1\left|W_{<1+\delta}\cap W_{>1-\varepsilon_{0},p}\right.\right) \le & \left(1-\dfrac{\theta}{h(n)\cdot\varepsilon_{0}{}^{3}}\right)^{pB}=\left(1-\dfrac{\theta/\varepsilon_{0}{}^{3}}{h(n)}\right)^{\frac{h(n)\cdot pn}{h^{2}(n)}}.\label{com24 item b part 2}
\end{aligned}
\end{equation}
As 
\[
\left(1-\dfrac{\theta/\varepsilon_{0}{}^{3}}{h(n)}\right)^{h(n)}\xrightarrow[n\to\infty]{}e^{-\theta/\varepsilon_{0}{}^{3}},\qquad\dfrac{pn}{h^{2}(n)}\xrightarrow[n\to\infty]{}\infty,
\]
 we obtain from (\ref{com24 item b part 1}) and (\ref{com24 item b part 2})
\begin{align}
\lim_{n\to\infty}P\left(U=1\left|W_{<1+\delta}\cap W_{>1-\varepsilon_{0},p}\right.\right) & =0.\label{com24 item b part 3}
\end{align}

Now we claim that the event $W_{<1+\delta}\cap\overline{W}_{>1-\varepsilon_{0},p}$
in the second term on the right-hand side of (\ref{com24 item b part 0})
is empty. In fact, the event $W_{<1+\delta}$ means that all sub-instances~$\mathcal{I}_{i}$ are of density less than $1+\delta$, and the event
$\overline{W}_{>1-\varepsilon_{0},p}$ means that most of them are
of density at most $1-\varepsilon_{0}.$ Since the overall density
is $\alpha>1-\varepsilon_{0}$, the two events do not meet for sufficiently
small $\delta,p$. Thus 
\begin{equation}
P\left(W_{<1+\delta}\cap\overline{W}_{>1-\varepsilon_{0},p}\right)=0.\label{com24 item b part 4}
\end{equation}

We turn to  the last term on the right-hand side of (\ref{com24 item b part 0}).
The condition $\overline{W}_{<1+\delta}$ implies that there is at
least one $1\le j\le B$ such that the density of $\mathcal{I}_{j}$
is at least $1+\delta$. Since the threshold of $2$-SAT is $1$,
this $\mathcal{I}_{j}$ is UNSAT w.h.p., and in particular $\mathcal{I}$
is such. Hence:
\begin{equation}
\lim_{n\to\infty}P\left(U=1\left|\overline{W}_{<1+\delta}\right.\right)=0.\label{com24 item b part 5}
\end{equation}

By (\ref{com24 item b part 0}), (\ref{com24 item b part 3}), (\ref{com24 item b part 4})
and (\ref{com24 item b part 5}), $\mathcal{I}$ is UNSAT w.h.p. 

\item{(b)} In this part we employ the approach of part (a) with minor changes.
We may assume $h(n)=\theta_{1}\sqrt{n}$ for some $\theta_{1}>0$. 

\begin{description}
\item{(i)}
Consider (\ref{com24 item b part 0}). In the
first term on the right-hand side, as $pn/h^{2}(n)\le\theta_{2}$
for some~$\theta_{2}>0$, by (\ref{com24 item b part 2}) we have
\begin{align}
\overline{\lim}_{n\to\infty}P\left(U=1\left|W_{<1+\delta}\cap W_{>1-\varepsilon_{0},p}\right.\right) & \le e^{-\theta\cdot\theta_{2}/\varepsilon_{0}{}^{3}}.\label{com24 item b part 3-1}
\end{align}
 (\ref{com24 item b part 4}) and (\ref{com24 item b part 5}) still
hold in this case. Thus, by (\ref{com24 item b part 0}), (\ref{com24 item b part 4}),
(\ref{com24 item b part 5}) and (\ref{com24 item b part 3-1}),
\[
\overline{\lim}_{n\to\infty}P(U=1)\le e^{-\theta\cdot\theta_{2}/\varepsilon_{0}{}^{3}}<1.
\]
In the other direction, let $\alpha^{'}$ be strictly between $\alpha$
and $1$. Similarly to (\ref{com24 item b part 0}),
\begin{align}
P\left(U=1\right) & \ge P\left(W_{<\alpha^{'}}\right)P\left(U=1\left|W_{<\alpha^{'}}\right.\right).\label{trm 3.5 part b label 0}
\end{align}
First, consider the second factor on the right-hand side of (\ref{trm 3.5 part b label 0}).
Given that $W_{<\alpha^{'}}$ has occurred, for some~$\left(m_{1},\ldots,m_{k}\right)\in I_{<\alpha^{'}}$
the event~$\bigcap_{i=1}^{B}\left\{ W_{i}=m_{i}\right\} $ has occurred.
Similarly to (\ref{com24 item b part 1}), 
\[
P\left(U=1\left|W_{<\alpha^{'}}\right.\right)\ge\prod_{i=1}^{B}P\left(U_{i}=1\left|W_{i}=\alpha^{'}\cdot h(n)\right.\right)=P\left(U_{1}=1\left|W_{1}=\alpha^{'}\cdot h(n)\right.\right)^{B}.
\]
By (\ref{eq:window 3}), for some $\theta_{3},\theta_{4}>0$ 
\begin{align*}
\begin{aligned}\underline{\lim}_{n\to\infty}P\left(U=1\left|W_{<\alpha^{'}}\right.\right) & \ge\lim_{n\to\infty}\left(1-\dfrac{\theta_{3}}{h(n)\cdot\left(1-\alpha^{'}\right){}^{3}}\right)^{n/h(n)}\\
 & =\lim_{n\to\infty}\left(1-\dfrac{\theta_{1}^{-1}\cdot\theta_{3}\cdot\left(1-\alpha^{'}\right)^{-3}}{\sqrt{n}}\right)^{\sqrt{n}/\theta_{1}}=e^{-\theta_{4}}>0.
\end{aligned}
\end{align*}
Now consider the first factor on the right-hand side of (\ref{trm 3.5 part b label 0}).
By (\ref{balls and bins 2}), w.h.p.\ the number of clauses in the
sub-instance~$\mathcal{I}_{i}$ with the maximal number of clauses
is bounded above by 
\[
\dfrac{m}{B}\cdot\left(1+o(1)\right)\cdot\alpha\cdot h(n).
\]
 Thus the density of all $\mathcal{I}_{j}$-s is bounded by 
\[
\dfrac{\left(1+o(1)\right)\cdot\alpha\cdot h(n)}{h(n)}\xrightarrow[n\to\infty]{}\alpha<\alpha^{'},
\]
namely
\begin{align}
P\left(W_{<\alpha^{'}}\right) & \xrightarrow[n\to\infty]{}1.\label{trm 3.5 part b label 2}
\end{align}
By (\ref{trm 3.5 part b label 0})-(\ref{trm 3.5 part b label 2})
\[
\underline{\lim}_{n\to\infty}P(U=1)\ge1\cdot e^{-\theta_{4}}>0.
\]
Therefore, $\mathcal{I}$ is SAT with probability bounded away from
both $0$ and $1$.

\item{(ii)} Similarly to (\ref{com24 item b part 0}), and with $p>0$ to
be determined later,
\begin{align}
P\left(U=1\right) & \le P\left(U=1\left|W_{\ge1,p}\right.\right)+P\left(\overline{W}_{\ge1,p}\right).\label{trm 3.5 part b ii  label 1}
\end{align}
Consider the first addend on the right-hand side of (\ref{trm 3.5 part b ii  label 1}).
Similarly to (\ref{com24 item b part 1}),
\[
P\left(U=1\left|W_{\ge1,p}\right.\right)\le\prod_{i=1}^{pB}P\left(U_{i}=1\left|W_{i}=h(n)\right.\right)=P\left(U_{1}=1\left|W_{1}=h(n)\right.\right)^{pB}.
\]
By (\ref{eq:window 1}), for some $0<\theta_{2}<1$ and $\theta_{3}>0$
\begin{align}
P\left(U=1\left|W_{\ge1,p}\right.\right) & \le\theta_{2}^{p\sqrt{n}/\theta_{3}}\xrightarrow[n\to\infty]{}0.\label{trm 3.5 part b ii  label 2}
\end{align}
Consider the second addend on the right-hand side of (\ref{trm 3.5 part b ii  label 1}).
Define the variables $X_{j}$,~$1\le j\le n$, as follows: $X_{j}=1$
if the $j$-th clause consists of variables from the first community,
and $X_{j}=0$ otherwise. Thus,~$X_{j}\sim\textmd{Ber}(1/B)$. The
variables $X_{1},\ldots,X_{n}$ are independent,~$\vert X_{j}\vert\le1$
for $1\le j\le n$ and 
\[
\sum_{j=1}^{n}V\left(X_{j}\right)=n\cdot\dfrac{1}{B}\left(1-\dfrac{1}{B}\right)=h(n)\left(1-\dfrac{h(n)}{n}\right)\xrightarrow[n\to\infty]{}\infty.
\]
Thus, by a version of the Central Limit Theorem \citep[ Corollary 2.7.1]{CLT stats}
\[
\dfrac{\sum_{j=1}^{n}X_{j}-E\left(\sum_{j=1}^{n}X_{j}\right)}{\sqrt{\sum_{j=1}^{n}V\left(X_{j}\right)}}\xrightarrow[n\to\infty]{\mathcal{D}}N(0,1).
\]
Clearly,~$E\left(\sum_{j=1}^{n}X_{j}\right)=E(W_{1})=h(n)$. Thus,
for large $n$ we have 
\begin{equation}
\begin{aligned}P(W_{1}\ge h(n)) & =P\left(\dfrac{W_{1}-h(n)}{\sqrt{h(n)(1-h(n)/n)}}\ge0\right)\approx\Phi(0)=\dfrac{1}{2}.\end{aligned}
\label{trm 3.5 part b ii  label 4}
\end{equation}
(We mention in passing that, in fact, we do not need the Central Limit
Theorem for our purpose. By \citep[Theorem 1]{CLT binomial p()greater than 0.25},
as~$W_{1}\sim B(n,h(n)/n)$ and $h(n)/n>1/n$ 
\[
\begin{aligned}P(W_{1}\ge h(n)) & =P(W_{1}\ge E(W_{1}))>\dfrac{1}{4}.\end{aligned}
\]
This inequality is weaker than (\ref{trm 3.5 part b ii  label 4}),
but suffices for the proof.)

Define the variables 
\begin{align*}
D_{i} & =\begin{cases}
1, & \qquad W_{i}\ge h(n),\\
0, & \qquad\textmd{otherwise,}
\end{cases}\qquad1\le i\le B.
\end{align*}
The $D_{i}$-s are $\textmd{Ber}(p_{0})$-distributed, where $p_{0}=P(W_{1}\ge h(n))$.
Let $D=\sum_{i=1}^{B}D_{i}$. By (\ref{trm 3.5 part b ii  label 4})
\[
\begin{aligned}E(D)=B\cdot P(W_{1}\ge h(n)) & >B/3.\end{aligned}
\]
Consider the proportion of sub-instances with at least $h(n)$ clauses.
We want to find a~$p>0$ such that~$\begin{aligned}P(D>pB) & \xrightarrow[n\to\infty]{}1.\end{aligned}
$ By \citep[Lemma 2]{Bins and balls dubashi}, the variables $D_{i}$
are negatively correlated, and hence 
\begin{align*}
V(D) & =\sum_{i=1}^{B}V(D_{i})+2\sum_{1\le i<j\le B}\textmd{Cov}(D_{i},D_{j})\\
 & \le B\cdot V(D_{1})=B\cdot p_{0}(1-p_{0})\le B/4.
\end{align*}
By the one-sided Chebyshev inequality for any $p_{1}>0$
\[
\begin{aligned}P \big( D-E\left( D\right)\ge-p_{1}B \big) & \ge1-\dfrac{V(D)}{V(D)+p_{1}^{2}B^{2}}\ge1-\dfrac{B/4}{p_{1}^{2}B^{2}}=1-\dfrac{1}{4p_{1}^{2}B}\xrightarrow[n\to\infty]{}1.\end{aligned}
\]
Thus 
\[
\begin{aligned}P\big(D\ge E(D)-p_{1}B\big) & =P\big(D\ge B/3-p_{1}B\big)\xrightarrow[n\to\infty]{}1.\end{aligned}
\]
Therefore for $p=1/6$ w.h.p.\ $D>B/6$. Thus 
\begin{equation}
P\left(\overline{W}_{\ge1,1/6}\right)\xrightarrow[n\to\infty]{}0.\label{trm 3.5 part b label 7}
\end{equation}

\item{(c)}
\begin{description}
\item{(i)}
Let $\alpha^{'}\in(1-\varepsilon_{0},\alpha)$. Similarly to (\ref{com24 item b part 0})
\begin{align}
P\left(U=0\right) & \le P\left(U=0\left|W_{<\alpha^{'}}\right.\right)+P\left(\overline{W}_{<\alpha^{'}}\right).\label{trm 3.5 part c label 1}
\end{align}
 Consider the first term on the right-hand side of (\ref{trm 3.5 part c label 1}).
Similarly to the proofs of the previous parts, given that the event
$W_{<\alpha^{'}}$ has occurred, the density of each sub-instance~$\mathcal{I}_{i}$ is less than~$\alpha^{'}$, and thus,
\begin{align*}
P\left(U_{i}=0\left|W_{<\alpha^{'}}\right.\right) & \le P\left(U_{i}=0\left|W_{i}=\alpha^{'}\cdot h(n)\right.\right),\qquad1\le i\le B.
\end{align*}
Employing the union bound 
\begin{equation}
\begin{aligned}
P\left(U=0\left|W_{<\alpha^{'}}\right.\right) & \le\sum_{i=1}^{B}P\left(U_{i}=0\left|W_{i}=\alpha^{'}\cdot h(n)\right.\right)\\ &=B\cdot P\left(U_{1}=0\left|W_{1}=\alpha^{'}\cdot h(n)\right.\right).
\end{aligned}
\label{trm 3.5 part c label 2}
\end{equation}
By (\ref{eq:window 3}), as $\alpha^{'}>1-\varepsilon_{0}$, for some
$\theta>0$ 
\begin{align}
P\left(U_{1}=0\left|W_{1}=\alpha^{'}\cdot h(n)\right.\right) & <\dfrac{\theta}{h(n)\cdot\left(1-\alpha^{'}\right)^{3}}.\label{trm 3.5 part c label 3}
\end{align}
By (\ref{trm 3.5 part c label 2}) and (\ref{trm 3.5 part c label 3}),
and as $h(n)=\omega(\sqrt{n})$
\begin{align}
P\left(U=0\left|W_{<\alpha^{'}}\right.\right) & \le\dfrac{n}{h^{2}(n)}\cdot\dfrac{\theta}{\left(1-\alpha^{'}\right)^{3}}\xrightarrow[n\to\infty]{}0.\label{trm 3.5 part c label 4}
\end{align}
By (\ref{trm 3.5 part c label 4}), (\ref{trm 3.5 part b label 2})
and (\ref{trm 3.5 part c label 1}),~$\mathcal{I}$ is SAT w.h.p

\item{(ii)}
Start from (\ref{trm 3.5 part b ii  label 1}). Consider the
first term on the right-hand side of (\ref{trm 3.5 part b ii  label 1}).
As~$h(n)=o(n)$, similarly to (\ref{trm 3.5 part b ii  label 2})
we have 
\begin{align*}
P\left(U=1\left|W_{\ge1,p}\right.\right) & \le\theta_{1}^{pn/h(n)}\xrightarrow[n\to\infty]{}0.
\end{align*}
By (\ref{trm 3.5 part b label 7}), for sufficiently small $p$, the
second term on the right-hand side of (\ref{trm 3.5 part b ii  label 1})
will vanish.Thus,~$\mathcal{I}$ is UNSAT w.h.p.
\end{description}
\item{(d)}
\begin{description}
\item{(i)}
In part (c).(i) we only used the fact that $h(n)=\omega(\sqrt{n})$,
so that the proof there applies here as well.

\item{(ii)}
In this case we may assume that $B=B_{0}$ is fixed. By (\ref{balls and bins 2}),
the density of the sub-instance~$\mathcal{I}_{i}$ with the maximal
number of clauses is bounded above by 
\[
\dfrac{1}{h(n)}\cdot\left(\dfrac{m}{B}+\sqrt{\dfrac{2m\ln B}{B}}\right)=1+\sqrt{\dfrac{2\ln B_{0}}{n/B_{0}}}.
\]
Thus, denoting $\alpha^{'}(n)=1+\sqrt{2B_{0}\ln B_{0}/n}$, we have
\begin{equation}
P\left(W_{\le\alpha^{'}(n)}\right)\xrightarrow[n\to\infty]{}1.\label{com24 item d part ii label 0}
\end{equation}
By (\ref{trm 3.5 part b label 0}),
\[
\begin{aligned}P\left(U=1\right)\ge P\left(W_{\le\alpha^{'}(n)}\right)P\left(U=1\left|W_{\le\alpha^{'}(n)}\right.\right) & .\end{aligned}
\]
By (\ref{eq:window 1}), and similarly to (\ref{com24 item b part 0}),
for some $\theta>0$,
\begin{equation}
\begin{aligned}P\left(U=1\left|W_{\le\alpha^{'}(n)}\right.\right) & \ge\theta^{B_{0}}>0.\end{aligned}
\label{com24 item d part ii label 2}
\end{equation}
Thus, by (\ref{com24 item d part ii label 0})-(\ref{com24 item d part ii label 2}),
$\mathcal{I}$ is SAT with probability bounded away from $0$. 

In the other direction, there is at least one sub-instance $\mathcal{I}_{i}$
with density at least $1$. Without loss of generality assume that
the density of the first sub-instance $\mathcal{I}_{1}$ is at least
$1$ and thus, for the same~$\theta$ as above
\[
P\left(U_{1}=1\right)\le\theta<1.
\]
Therefore,
\[
\begin{aligned}P\left(U=1\right)=\prod_{i=1}^{B}P\left(U_{i}=1\right) & \le P\left(U_{1}=1\right)<1.\end{aligned}
\]
Thus, $\mathcal{I}$ is SAT with probability bounded away from both
$0$ and $1$.
\end{description}
\end{description}

\end{description}
\qed \end{Proof}

The proof of Theorem \ref{thm3.9} follows Chvátal and Reed \citep{threshold of 2 item ca}. The case $p=0$ follows from Proposition \ref{prop3.2}. We will thus assume that $p>0$.

We first recall two definitions and their relevance to the satisfiability/unsatisfiability of an instance.

\begin{definition}\citep{threshold of 2 item ca}\label{def-of-bicycle} A \emph{bicycle} is a formula that consists of at least two distinct variables $v_1,\ldots, v_s$ and clauses $C_0,C_1,\ldots,C_s$ with the following structure: there are literals $l_1,\ldots,l_s$ such that each $l_i$ is either $v_i$ or $\overline{v}_i$, we have $C_i=\overline{l}_i\bigvee l_{i+1}$ for all $1\leq i\leq s-1$, $C_0=u\bigvee l_1$, and $C_s=\overline{l}_s\bigvee v$ where $u,v\in\{v_1,\ldots,v_s,\overline{v}_1,\ldots,\overline{v}_s\}$.
\end{definition}

Chvátal and Reed \citep{threshold of 2 item ca} proved that every unsatisfiable formula contains a bicycle.

\begin{definition}\citep{threshold of 2 item ca}
A \emph{snake} is a sequence of distinct literals $l_1,\ldots,l_s$ such that no $l_i$ is the complement of another.
\end{definition}

Chvátal and Reed \citep{threshold of 2 item ca} proved that, for a snake $A$ consisting of the literals $l_1,\ldots,l_s$, the formula $F_A$, consisting of the $s+1$ clauses $\overline{l}_i\bigvee l_{i+1}$ for all $0\leq i\leq s$ with $l_0=l_{s+1}=\overline{l}_t$, is unsatisfiable.

\vspace{10pt}
\begin{Proof}{\ of Theorem \ref{thm3.9}}
Suppose that
\begin{equation*}
\lim_{n\rightarrow\infty}\frac{m}{n}=r.
\end{equation*}
We have to show that for $r<1$ our formula is satisfiable w.h.p., while for $r>1$ it is unsatisfiable w.h.p.

First suppose that $r<1$. Let $p'$ be the probability that our formula contains a bicycle. We will derive an upper bound for $p'$. To derive this upper bound, we will add up the probabilities of our formula containing each specific bicycle. Thus, first take some specific bicycle, consisting of $s+1$ clauses $C_0,C_1,\ldots,C_s$ as in Definition \ref{def-of-bicycle} for some $2\leq s\leq n$. Also, suppose that exactly $j$ out of the clauses $C_1,C_2,\ldots,C_{s-1}$ consist of two variables from the same community. There are at most $m^{s+1}$ choices as to which of the $m$ clauses will make up the clauses $C_0,C_1,C_2,\ldots,C_s$ in our formula. The probability of a clause in the formula to be some specific clause, with both variables in the same community, in our bicycle is 
\begin{equation*}
\frac{1-p}{4B\binom{n/B}{2}},
\end{equation*}
whereas if the variables are in different communities then this probability is
\begin{equation*}
\frac{p}{\binom{B}{2}\left(\frac{2n}{B}\right)^2}.
\end{equation*}
Then the probability that our formula will contain this specific bicycle is bounded above by
\begin{equation*}
\frac{(1-p)^jp^{s-1-j}m^{s+1}}{\left(\binom{2n}{2}\right)^2\left(4B\binom{n/B}{2}\right)^j\left(\binom{B}{2}\left(\frac{2n}{B}\right)^2\right)^{s-1-j}}=\frac{B^{s-1}(1-p)^jp^{s-1-j}m^{s+1}}{n^{2s-j}(2n-1)^2(2n-2B)^j(2(B-1))^{s-1-j}}.
\end{equation*}
Now we count the number of possible bicycles. Suppose we restrict ourselves to bicycles such that exactly $j$ clauses out of $C_1,C_2,\ldots,C_{s-1}$ as defined above each consists of two variables from the same community. There are $\binom{s-1}{j}$ ways to choose these clauses. Then if we pick the literals for the bicycle one at a time, we have $2n$ choices for the first literal since we have $n$ Boolean variables in total. If $C_1$ is supposed to contain both variables from the same community, then there are $\frac{2n}{B}-2$ choices for the second literal. On the other hand, if $C_1$ is supposed to contain variables from different communities, then there are $\frac{2(B-1)n}{B}$ choices for the second literal. Continuing in this way, we see that there are less than $2n\left(\frac{2n}{B}\right)^j\left(\frac{2(B-1)n}{B}\right)^{s-1-j}$ choices for the literals in the bicycle. Also, there are at most $s^2$ choices for $u$ and $v$. Hence, assuming $2n>B$, we have
\begin{align*}
p'&<\sum_{s=2}^{n}s^2\sum_{j=0}^{s-1}\binom{s-1}{j}(2n)^s\left(\frac{1}{B}\right)^j\left(\frac{B-1}{B}\right)^{s-1-j}\cdot\frac{B^{s-1}(1-p)^jp^{s-1-j}m^{s+1}}{n^{2s-j}(2n-1)^2(2n-2B)^j(2(B-1))^{s-1-j}}\\
&=\sum_{s=2}^{n}s^2\sum_{j=0}^{s-1}\binom{s-1}{j}\frac{2^{j+1}n^{j-s}(1-p)^jp^{s-1-j}m^{s+1}}{(2n-1)^2(2n-2B)^j}\\
&=\sum_{s=2}^n\frac{2s^2p^{s-1}m^{s+1}}{n^s(2n-1)^2}\sum_{j=0}^{s-1}\binom{s-1}{j}\left(\frac{2n(1-p)}{p(2n-2B)}\right)^j\\
&=\frac{2m^2}{n(2n-1)^2}\sum_{s=2}^ns^2\left(\frac{2nm-2pBm}{n(2n-2B)}\right)^{s-1}.
\end{align*}
By a geometric series argument, the sum above is finite, and so $p'=O\left(\frac{1}{n}\right)$. Thus, the satisfiability threshold is at least $1$.

Now suppose $r>1$. We will also assume that $p<1$. (If $p=1$, the proof is actually simpler.) For each $n\in\mathbb{N}$, choose a $t=t(n)\in\mathbb{N}$ in such a way that
\begin{equation}\label{eqn3}
\lim_{n\rightarrow\infty}t/\log n=\infty,\qquad\lim_{n\rightarrow\infty}t/n^{1/9}=0.
\end{equation}
Let $s=2t-1$. We will show that our formula contains a formula $F_A$ of a snake $A$ consisting of $s$ literals w.h.p. Thus, our formula will be unsatisfiable w.h.p. We use the second moment method. Let $X=\sum X_A$, where $X_A=1$ if our formula contains each clause of $F_A$ exactly once, and $X_A=0$ otherwise. We will prove that
\begin{equation}\label{eqn7}
E(X^2)\leq(1+o(1))E(X)^2,
\end{equation}
from which the desired result may be deduced using Chebyshev's inequality. Consider an arbitrary snake $A$. Suppose that $F_A$ contains exactly $t_1$ clauses each consisting of a pair of variables from different communities and exactly $t_2$ clauses each consisting of variables from the same community. We have $E(X_A)=f(t_1,t_2)$, where
\begin{align*}
f(x_1,x_2)&=\sum_{i=x_1}^{m-x_2}p^i(1-p)^{m-i}\binom{m}{i}\binom{i}{x_1}\binom{m-i}{x_2}x_1!x_2!\left(\frac{1}{4\binom{B}{2}\frac{n^2}{B^2}}\right)^{x_1}\left(\frac{1}{4B\binom{n/B}{2}}\right)^{x_2}\\
&\qquad\qquad\cdot\left(1-\frac{x_1}{4\binom{B}{2}\frac{n^2}{B^2}}\right)^{i-x_1}\left(1-\frac{x_2}{4B\binom{n/B}{2}}\right)^{m-i-x_2}\\
&=(1-p)^{x_2}p^{x_1}\left(\frac{1}{4\binom{B}{2}\frac{n^2}{B^2}}\right)^{x_1}\left(\frac{1}{4B\binom{n/B}{2}}\right)^{x_2}\frac{m!}{\left(m-x_1-x_2\right)!}\\
&\quad\cdot\left((1-p)\left(1-\frac{x_2}{4B\binom{n/B}{2}}\right)+p\left(1-\frac{x_1}{4\binom{B}{2}\frac{n^2}{B^2}}\right)\right)^{m-x_1-x_2}.
\end{align*}
Take two snakes $A$ and $A'$, where $F_A$ contains exactly $t_1$ clauses with variables from different communities and exactly $t_2$ clauses with variables from the same community, and $F_{A'}$ contains exactly $t_3$ clauses with variables from different communities and exactly $t_4$ clauses with variables from the same community. Also, suppose $F_A$ and $F_{A'}$ share precisely $i_1$ clauses with variables in different communities and precisely $i_2$ clauses with variables from the same community. Then $E(X_A X_{A'})=f(t_1+t_3-i_1,t_2+t_4-i_2)$. Since $m=O(n)$, we have
\begin{equation}\label{eqn4}
f(x_1,x_2)=(1+o(1))\left(\frac{Bpm}{2(B-1)n^2}\right)^{x_1}\left(\frac{B(1-p)m}{2n^2}\right)^{x_2}
\end{equation}
uniformly in both cases if we assume that $x_1,x_2=O(n^{\alpha})$ where $\alpha<1/2$.

Now let us count the snakes $A$ such that $F_A$ contains exactly $t_1$ clauses with variables from different communities and exactly $t_2$ clauses with variables from the same community. We denote the set of all such snakes as $S_{t_1,t_2}$. First, we may view $F_A$ as a directed graph with vertices $y_1,\ldots, y_s$ (where each $y_i$ is the variable such that $l_i$ is $y_i$ or $\overline{y}_i$) and edges $y_iy_{i+1}$, $0\leq i\leq s$, with $y_0=y_{s+1}=y_t$. This directed graph consists of two direct cycle graphs, each consisting of $t$ vertices and having exactly one vertex in common (the vertex $y_0=y_{s+1}=y_t$). Each edge corresponds to a clause in $F_A$. Consider the $t_2$ edges corresponding to the $t_2$ clauses with variables in different communities. Let $j_1$ and $j_2$ be the number of such edges in each of the two cycle graphs that make up the whole graph. We can see that $j_1,j_2\neq 1$. For $k=1,2$ for the cycle with the $j_k$ edges, there will be $\binom{t}{j_k}$ ways to choose these $j_k$ edges. These $j_k$ edges will then partition the set of vertices into $\max\{1,j_k\}$ groups, where the variables corresponding to the vertices in a group will belong to the same community. Thus, the number of ways of choosing the community of each of the variables corresponding to the vertices in this cycle graph is the chromatic number of the cycle graph consisting of $\max\{1,j_k\}$ vertices in $B$ colours or $(B-1)^{j_k}+\left(j_k-1\right)(-1)^{j_k}$. After choosing all of these communities for each cycle graph, we are left with choosing the variables from these communities, and there will be at least $\frac{n}{B}-s$ choices per vertex after making the choice for the $v_t$ variable. Putting it altogether, the number of such snakes will be bounded below by
\begin{equation*}
\frac{2n}{B}\left(\frac{2n}{B}-2s\right)^{s-1}\sum_{\substack{j_1+j_2=t_1\\j_1\neq 1,j_2\neq 1}}\binom{t}{j_1}\binom{t}{j_2}\cdot\frac{\left((B-1)^{j_1}+(B-1)(-1)^{j_1}\right)\left((B-1)^{j_2}+(B-1)(-1)^{j_2}\right)}{B}
\end{equation*}
and bounded above by
\begin{equation*}
\left(\frac{2n}{B}\right)^s\sum_{\substack{j_1+j_2=t_1\\j_1\neq 1,j_2\neq 1}}\binom{t}{j_1}\binom{t}{j_2}\cdot\frac{\left((B-1)^{j_1}+(B-1)(-1)^{j_1}\right)\left((B-1)^{j_2}+(B-1)(-1)^{j_2}\right)}{B}.
\end{equation*}
By \eqref{eqn3}, the latter is asymptotic to the actual number of such snakes as $n\rightarrow\infty$. By \eqref{eqn4}, we thus have:
\begin{align}
E(X)&\sim\sum_{t_1=0}^{2t}\left(\frac{Bpm}{2(B-1)n^2}\right)^{t_1}\left(\frac{B(1-p)m}{2n^2}\right)^{2t-t_1}\left(\frac{2n}{B}\right)^{2t-1}\nonumber\\
&\qquad\qquad\cdot\sum_{j_1+j_2=t_1}\binom{t}{j_1}\binom{t}{j_2}\cdot\frac{\left((B-1)^{j_1}+(B-1)(-1)^{j_1}\right)\left((B-1)^{j_2}+(B-1)(-1)^{j_2}\right)}{B}\nonumber\\
&=\frac{1}{B}\left(\frac{2n}{B}\right)^{2t-1}\left(\frac{B(1-p)m}{2n^2}\right)^{2t}\left(\sum_{j=0}^t\binom{t}{j}\left(\frac{p}{(B-1)(1-p)}\right)^j\left((B-1)^j+(B-1)(-1)^j\right)\right)^2\nonumber\\
&=\frac{1}{B}\left(\frac{2n}{B}\right)^{2t-1}\left(\frac{B(1-p)m}{2n^2}\right)^{2t}\left(\left(1+\frac{p}{(1-p)}\right)^t+(B-1)\left(1-\frac{p}{(B-1)(1-p)}\right)^t\right)^2\nonumber\\
&=\frac{1}{2n}\left(\left(\frac{m}{n}\right)^t+(B-1)\left(\frac{(1-p)m}{n}-\frac{mp}{n(B-1)}\right)^t\right)^2\nonumber\\
&\sim\frac{1}{2n}\left(\frac{m}{n}\right)^{2t}\label{eqn8}.
\end{align}
By \eqref{eqn3}, we have
\begin{equation*}
E(X_AX_{A'})=(1+o(1))E(X_A)E(X_{A'})\left(\frac{2(B-1)n^2}{Bpm}\right)^{i_1}\left(\frac{2n^2}{B(1-p)m}\right)^{i_2}
\end{equation*}
uniformly in the range $0\leq i_1,i_2\leq 2t$. In particular, if $F_A$ and $F_{A'}$ have no clauses in common, then $E(X_AX_{A'})=(1+o(1))E(X_A)E(X_{A'})$. Thus, to prove \eqref{eqn7}, our main concern will be when $F_A$ and $F_{A'}$ have clauses in common. To deal with this case, we will derive an upper bound for 
\begin{equation*}
\sum_{|F_A\cap F_{A'}|=i} E\left(X_AX_{A'}\right)
\end{equation*}
(where $F_A\cap F_{A'}$ denotes the set of common clauses of $F_A$ and $F_{A'}$) for each $1\leq i\leq 2t$. First consider how we can construct two snakes $A$ and $A'$ such that $F_A$ and $F_{A'}$ have $i$ clauses in common and account for its contribution to the above sum. Viewing $F_A$ and $F_{A'}$ as graphs as above, we let $F_{AA'}$ be their intersection, with isolated vertices removed. Suppose that $F_{AA'}$ contains $i$ edges and $j$ vertices. To construct the possible snakes $A$ and $A'$, we create a procedure with five steps:

1) Choose $j$ terms of $A$ for membership in $F_{AA'}$.

2) Assign variables to these $j$ terms.

3) Choose which positions in the snake $A'$ will be filled with terms in $F_{A{A'}}$. 

4) Assign variables to the positions in ${A'}$ picked out in step 2).

5) Assign variables to all other positions in $A$ and ${A'}$. 

For 1), we can select our $j$ terms of $A$ as follows. We first decide if the edge $y_0y_1$ is in $F_{A{A'}}$ or not, and then, for each $1\leq r\leq s$, we place a marker at $y_r$ if exactly one of $y_{r-1}y_r$ and $y_ry_{r+1}$ is in $F_{A{A'}}$. The total number of markers will be between $2(j-i)-1$ and $2(j-i)+2$, and so the total number of choices for the $j$ terms is at most $2\binom{s+3}{2j-2i+2}$. Thus, the total number of choices for 3) will also be at most $2\binom{s+3}{2j-2i+2}$. Also, we have at most $tk!2^k$ choices for step 4), where $k$ is the number of components in $F_{A{A'}}$. 

For step 2), if we impose the restriction that $i_1$ edges among the $j$ vertices correspond to the clauses with variables in different communities, then the number of ways to assign such variables is $\binom{i}{i_1}\left(\frac{2n}{B}\right)^jB^k(B-1)^{i_1}$. As well, for step 5), if we impose the restrictions that, of the remaining $2t-i$ clauses in $A$, there are exactly $t_1$ with variables in different communities, and that of the remaining $2t-i$ clauses in ${A'}$ there are exactly $t_2$ clauses with variables in different communities, then the number of ways to assign such variables is bounded above by $\binom{2t-i}{t_1}\binom{2t-i}{t_2}\left(\frac{2n}{B}\right)^{2s-2j}(B-1)^{t_1+t_2}$.

First suppose that $1\leq i\leq t-1$. Then none of the components of $F_{A{A'}}$ may contain loops, so that $k=j-i$. Putting it all together, weighing all of the possible pairs of snakes $A$ and ${A'}$, appropriately using \eqref{eqn3}, we obtain
\begin{align*}
\sum_{|F_A\cap F_{A'}|=i}E\left(X_AX_{A'}\right)&<\sum_{j\geq
i+1}\frac{9}{2}\binom{s+3}{2j-2i+2}^2t\cdot(j-i)!(2B)^{j-i}\\
&\qquad\qquad\cdot\left(\left(\frac{2n}{B}\right)^j\sum_{i_1=0}^i\binom{i}{i_1}\left(\frac{Bpm}{2n^2}\right)^{i_1}\left(\frac{B(1-p)m}{2n^2}\right)^{i-i_1}\right)\\
&\qquad\qquad\cdot\left(\left(\frac{2n}{B}\right)^{s-j}\sum_{t_1=0}^{2t-i}\binom{2t-i}{t_1}\left(\frac{Bpm}{2n^2}\right)^{t_1}\left(\frac{B(1-p)m}{2n^2}\right)^{2t-i-t_1}\right)^2\\
&<\frac{9B^2t(2t+2)^4}{8n^2}\left(\frac{m}{n}\right)^{4t-i}\sum_{j\geq i+1}\left(\frac{B^3(2t+2)^4}{n}\right)^{j-i}
\end{align*}
for sufficiently large $n$. Thus by \eqref{eqn8} we have for sufficiently large n
\begin{align*}
\frac{\sum_{|F_A\cap F_{A'}|=i}E\left(X_AX_{A'}\right)}{E(X)^2}&<5t(2t+2)^4\left(\frac{n}{m}\right)^i\sum_{j\geq i+1}\left(\frac{B^3(2t+2)^4}{n}\right)^{j-i}\\
&<\frac{2600B^3t^9}{n}\left(\frac{n}{m}\right)^i.
\end{align*}
Now suppose that $t\leq i\leq 2t$. We have two possibilities for the components of $F_{A{A'}}$. Either none of them contains loops or exactly one of them contains a loop and the number of loops in this component is exactly $1$ or $2$, where the possible loops are $y_0,y_1,\ldots,y_t$ and $y_t,y_{t+1}\ldots,y_{s+1}$. In either case we have $k\leq j-i+2$. Thus,
\begin{align*}
\sum_{|F_A\cap F_{A'}|=i}E\left(X_AX_{A'}\right)&<\sum_{j\geq i+1}\frac{9}{2}\binom{s+3}{2j-2i+2}^2t\cdot(j-i)!(2B)^{j-i}B^2\\
&\qquad\qquad\cdot\left(\left(\frac{2n}{B}\right)^j\sum_{i_1=0}^i\binom{i}{i_1}\left(\frac{Bpm}{2n^2}\right)^{i_1}\left(\frac{B(1-p)m}{2n^2}\right)^{i-i_1}\right)\\
&\qquad\qquad\cdot\left(\left(\frac{2n}{B}\right)^{s-j}\sum_{t_1=0}^{2t-i}\binom{2t-i}{t_1}\left(\frac{Bpm}{2n^2}\right)^{t_1}\left(\frac{B(1-p)m}{2n^2}\right)^{2t-i-t_1}\right)^2\\
&<\frac{9B^6t(2t+2)^4}{2n^2}\left(\frac{m}{n}\right)^{4t-i}\sum_{j\geq i+1}\left(\frac{B^3(2t+2)^4}{n}\right)^{j-i}.
\end{align*}
By \eqref{eqn8}, for sufficiently large $n$
\begin{align*}
\frac{\sum_{|F_A\cap F_{A'}|=i}E\left(X_AX_{A'}\right)}{E(X)^2}&<20B^4t(2t+2)^4\left(\frac{n}{m}\right)^i\sum_{j\geq i+1}\left(\frac{B^3(2t+2)^4}{n}\right)^{j-i}\\
&<\frac{10400B^7t^9}{n}\left(\frac{n}{m}\right)^i.
\end{align*}
Thus
\begin{equation*}
\sum_{i=1}^{2t}\frac{\sum_{|F_A\cap F_{A'}|=i} E\left(X_AX_{A'}\right)}{E(X)^2}<\sum_{i=1}^{2t}\frac{10400B^7t^9}{n}\left(\frac{n}{m}\right)^i=o(1),
\end{equation*}
from which we can deduce \eqref{eqn7}.

\qed
\end{Proof}

\section{Empirical results}

To test the question posed after Theorem \ref{thm3.9}, we have conducted the following experiment. We have taken $n=10^6$, and~$m=n+c\cdot n^{2/3},$ with
$c=-1,0,1,2$. (This non-symmetric range was due to preliminary simulations, that showed that the interesting window is actually centered somewhat above $n$. For each such ~$m$, we generated $10^5$ random instances from  $F\left(n,m,2,P_{(1,1)}\right)$ and 
$F\left(n,m,1,P_{(2)}\right)$ (which
is just the random model), tested each instance using the SAT solver SAT4J, described in \citep{sat4j}, and calculated the percentage of satisfiable instances in each group. To complete the picture, we did the same for the model~$F\left(n,m,2,P_{(2)}\right)$.

The results are presented in Table~\ref{table4}. The first two models show remarkably similar results. Unsurprisingly, the third model leads to lower satisfiability probabilities.
\begin{centering}
\setcellgapes{7pt}
\begin{table}[ht]
\makegapedcells
\centering
\begin{tabular}{!{\VRule[2pt]}c!{\VRule}c!{\VRule}c!{\VRule}c!{\VRule}c!{\VRule[2pt]}}
\specialrule{2pt}{0pt}{0pt}
\multicolumn{1}{!{\VRule[2pt]}c!{\VRule}}{\diaghead(-4,1){aaaaaaaaaaaaaaaaa}{\boldmath{$F$}}{\boldmath{$m$}}}&
$0.99\cdot 10^6$ & $\quad 10^6\quad $ & $1.01\cdot 10^6$ & $1.02\cdot 10^6$ \\ \hline
& & & &\\[-2.6ex] \hline
$F\left(n,m,2,P_{(1,1)}\right)$ & $0.980$  & $0.909$ & $0.641$ & $0.201$\\\hline
$F\left(n,m,1,P_{(2)}\right)$ & $\:\:\:0.980\:\:\:$  & $0.908$ & $0.644$ & $0.203$ \\\hline
$F\left(n,m,2,P_{(2)}\right)$ & $0.946$  & $0.827$ & $0.521$ & $0.142$ \\
\specialrule{2pt}{0pt}{0pt}
\end{tabular}
\caption{Percentage of satisfiable instances (out of $10^5$ instances) for $n=10^6$.}
\label{table4}
\end{table} 
\end{centering}


\section{Conclusions}

We have dealt with the satisfiability threshold of a particular model of SAT. This model highlights one of the features in which so-called community-structured SAT instances differ from classical SAT instances. 
Namely, the set of variables decompose into several disjoint subsets-communities. The significance of these communities stems from the fact that clauses tend to contain variables from the same community. We have shown, roughly speaking, that the satisfiability threshold of such instances tends to be lower than for regular instances. Moreover, if the communities are very small, the threshold may even vanish.

The paper leaves a lot to study for industrial SAT instances. 
To begin with, there are other features considered in the literature as being characteristic of industrial instances. For example, in the scale-free structure, the variables
are selected by some heavy-tailed distribution. Moreover, even regarding the issue of communities, there is more to be done. We assumed here that all communities are of the same size. Obviously, there is no justification for this assumption beyond the fact that it simplifies significantly the analysis of the model. What can be said about the threshold if there are both small and large communities? Even prior to that, what would be reasonable to assume regarding the probability of a variable to be selected from each of the communities?

\section*{Acknowledgments}

We would like to express our gratitude to David Wilson for helpful
information regarding the topic of this paper, and to the two anonymous referees for their comments on the first version of the paper.

\newpage
\bibliographystyle{named}
\bibliography{ijcai18}

\begin{thebibliography}{9}
\bibitem[Achlioptas and Peres, 2004]{threshold achiloptas} Dimitris Achlioptas
and Yuval Peres, ``The threshold for random $k$-SAT is $2^{k}\ln2-O(k)$'',
\\Journal of the American Mathematical Society 17.4 (2004), 947\textendash{}973.

\bibitem[Adams and  Guillemin, 1996]{AdamsGuillemin} Malcolm Adams,  and Victor Guillemin, ``{Measure theory and probability}'', {Corrected reprint of the 1986 original}, {Birkh\"{a}user Boston, Inc., Boston, MA}, (1996).

\bibitem[Alon and Lubetzky, 2009]{alon}Noga Alon and Eyal Lubetzky, ``Poisson approximation for non-backtracking random walks'',\\Israel Journal of Mathematics 174.1 (2009), 227\textendash{}252.

\bibitem[Ans{\'o}tegui et al., 2015]{Jordy 2015}Carlos Ans{\'{o}}tegui, Maria L. Bonet, Jes{\'{u}}s Gir{\'{a}}ldez{-}Cru, and
Jordi Levy, ``On the Classification of Industrial {SAT} Families'', \textit{Artificial Intelligence Research and Development - Proc. of the 18th International Conference of the Catalan Association for Artificial Intelligence}, Valencia, Catalonia, Spain, October 21--23 (2015), 163--172.

\bibitem[Ans{\'o}tegui et al., 2009A]{Jordy 2009 B1}Carlos Ansótegui,
Mar\'ia L. Bonet, and Jordi Levy, ``Towards industrial-like random SAT instances'', 
 \textit{Proc. of the 21st International Joint Conference on Artificial Intelligence, IJCAI 2009} (2009).

\bibitem[Ans{\'o}tegui et al., 2009B]{Jordy 2009B}Carlos Ans{\'o}tegui, Mar{\'i}a L. Bonet,
and Jordi Levy, ``On the Structure of Industrial SAT Instances'', 
\textit{Principles and Practice of Constraint Programming - CP 2009}, Lecture Notes in Comput. Sci., 5732,
I. P. Gent, Ed., 
Springer Berlin Heidelberg (2009), 127--141.

\bibitem[Ansótegui et al., 2012]{Jordy 2009C}Carlos Ansótegui,
Jesús Giráldez-Cru, and Jordi Levy, ``The community structure of SAT formulas'', 
\textit{Proc. of Theory and Applications of Satisfiability Testing\textendash{}SAT
2012: 15th International Conference}, Trento, Italy, June 17--20,
2012, A. Cimatti, R. Sebastiani, Eds., 410--423.

\bibitem{sat4j}Daniel Le Berre and Anne Parrain, ``The Sat4j library,
release 2.2, system description'', Journal on Satisfiability,
Boolean Modeling and Computation 7 (2010), 59\textendash{}64.

\bibitem[Bollobás et al., 2001]{scaling window}Béla
Bollobás, Christian Borgs, Jennifer T. Chayes, Jeong H. Bim, and David B. Wilson, ``The
scaling window of the $2$-SAT transition'', Random Structures
\& Algorithms 18 (2001), 3, 201\textendash{}256. 

\bibitem[Chvátal and Reed, 1992]{threshold of 2 item ca}Vaclav Chvátal
and B. Reed, ``Mick gets some (the odds are on his side)'', \textit{Proc. 33rd Symp. Foundations of Computer Science} (1992), 620\textendash{}627. DINA : CHECK REEDS FIRST NAME

\bibitem[Cook, 1971]{cook}Stephen A. Cook, ``The complexity of theorem proving procedures'', 
\textit{Proc. of the Third Annual ACM STOC} (1971), 151\textendash{}158.


\bibitem[Coja-Oghlan and Panagiotou, 2016]{thresh coja 2016}Amin Coja-Oghlan and Konstantinos Panagiotou, ``The
asymptotic $k$-SAT threshold'', Advances in Mathematics 288 (2016), 985--1068.

\bibitem[Ding et al., 2015]{large k}Jian Ding, Allan Sly, and Nike Sun, \textquotedblleft Proof of the satisfiability conjecture for large $k$\textquotedblright , (English summary) \textit{STOC'15\textemdash Proc.
of the 2015 ACM Symposium on Theory of Computing} (2015), 59--68.

\bibitem[Díaz et al., 2009]{Diaz}Josep Díaz, Lefteris Kirousis, Dieter Mitsche, and Xavier Pérez-Giménez, ``On the satisfiability threshold of formulas with three literals per clause'', Theoretical Computer Science 410 (2009) 30, 2920--2934.

\bibitem{Bins and balls dubashi}Devdatt Dubhashi and Desh Ranjan, ``Balls
and bins: a study in negative dependence", Random
Structures \& Algorithms 13 (1998) 2, 99--124.

\bibitem[Fernandez de la Vega, 1992]{threshold of 2 item Fe}Wenceslas Fernandez
de la Vega, On random $2$-SAT, unpublished manuscript (1992).

\bibitem[Franco and Paull, 1983]{threshold franco}John V. Franco and Marvin C.
Paull, ``Probabilistic analysis of the Davis\textendash{}Putnam
procedure for solving the satisfiability problem'', Discrete Applied
Mathematics 5(1) (1983), 77--87.

\bibitem[Giráldez-Cru and Levy, 2015]{Jordy 2015 modularity}Jesús Giráldez-Cru
and Jordi Levy, ``A modularity-based random SAT instances generator'', 
\textit{Proc. of the 24th International Joint Conference on Artificial
Intelligence, IJCAI\textquoteright 15} (2015), 1952\textendash{}1958.

\bibitem[Giráldez-Cru and Levy, 2016]{Jordy 2016}Jesús Giráldez-Cru and Jordi Levy, ``Generating SAT
instances with community structure'', Artificial Intelligence
238 (2016), 119\textendash{}134.

\bibitem[Giráldez-Cru and Levy, 2017]{Levy17}Jesús
Giráldez-Cru and Jordi Levy, ``Locality in Random SAT Instances'', \textit{Proc. of the 26th International Joint Conference on Artificial Intelligence,
IJCAI\textquoteright 17} (2017) 638\textendash{}644.

\bibitem[Greenberg and Mohri, 2014]{CLT binomial p()greater than 0.25} Spencer Greenberg and Mehryar Mohri, 
``Tight lower bound on the probability of a binomial exceeding its expectation'', Statistics \& Probability Letters 86 (2014), 91\textendash{}98.

\bibitem[Goerdt, 1992]{threshold of 2 item G1}Andreas Goerdt, ``A threshold
for unsatisfiability'',  \textit{Proc. Mathematical Foundations of Computer
Science 1992: 17th International Symp.}, Prague, Czechoslovakia, August
24\textendash{}28, 1992, 629 Springer, Berlin (1992), 264\textendash{}274.

\bibitem[Hajiaghayi and Sorkin, 2003]{Hajiaghayi}M. Hajiaghayi ans G. B. Sorkin, ``The satisfiability threshold of random 3-SAT is at least 3.52'', Research Report RC22942, IBM, October 2003. 

\bibitem[Kaporis et al., 2003]{Sat3LowerBound2}
Alexis C. Kaporis, Lefteris M. Kirousis, and Efthimios G. Lalas. ``The Probabilistic Analysis of a Greedy Satisfiability Algorithm'', Random Structures \& Algorithms 28(4) (2006), 444--480. 



\bibitem[Kirousis et al., 1998]{thresh kkk89}Lefteris M. Kirousis, Evangelos Konstantinou Kranakis, Danny Krizanc, and
Y. C. Stamatiou, ``Approximating the unsatisfiability threshold of  random formulas'', Random Structures \& Algorithms 12(3) (1998), 253--269.

\bibitem[Kolchin et al., 1978]{allocation}Valentin  F. Kolchin, Boris  A. Sevast\' yanov, and
Vladimir P. Chistyakov, ``Random allocations'', V. H. Winston \& Sons, Washington, D.C., (1978).

\bibitem{CLT stats} Eric Leo Lehmann, ``Elements of large-sample theory'', Springer-Verlag, New York, (1999).


\bibitem[Liang et al., 2018]{Oh2018}Jia Hui Liang, Chanseok Oh,  Minu Mathew,   Ciza Thomas, Chunxiao Li, and Vijay Ganesh, 
``{Machine learning-based restart policy for {CDCL} {SAT}  solvers}'',
\textit{Theory and applications of satisfiability testing -- {SAT} 2018}, {Lecture Notes in Comput. Sci.}, {10929}, {Springer, Cham} (2018), {94--110}.

\bibitem[Mertens et al., 2006]{thresh zecchina 1}Stephan
Mertens, Marc Mézard, and Riccardo Zecchina, ``Threshold values of random $k$-SAT from the cavity method'', Random Structures \& Algorithms 28 (2006), 340--373. 

\bibitem[Mézard and Zecchina, 2002]{thresh zecchina 2}Marc Mézard and Riccardo Zecchina,  ``Random $k$-satisfiability
problem: From analytic solution to an efficient algorithm'', Physical Review E 66 (2002), 056126.

\bibitem[Newman, 2006]{modularity newman A}Mark E. J. Newman, ``Finding community structure in networks using the eigenvectors of matrices'', Physical Review E74(3) (2006), 036104. 

\bibitem[Newsham et al., 2014]{newsham}Zack Newsham, Vijay Ganesh, Sebastian Fischmeister, Gilles Audemard, and Laurent Simon, ``Impact of community structure on {SAT} solver performance", \textit{Theory and applications of satisfiability testing -- SAT 2014}, Lecture Notes in Comput. Sci., 8561, Springer (2014), 252--268.

\bibitem[Oh, 2015]{Oh2015}Chanseok Oh, ``{Between {SAT} and {UNSAT}: the fundamental difference in {CDCL} {SAT}}, \textit{Theory and applications of satisfiability testing -- {SAT} 2015}, Lecture Notes in Comput. Sci., 9340, Springer, Cham (2015), 307--323.

\bibitem[Oh, 2016]{Oh2016}Chanseok Oh, ``Improving SAT solvers by exploiting empirical characteristics of CDCL'', PHD thesis (Doctoral dissertation), New York University, \\ https://cs.nyu.edu/media/publications/oh\_chanseok.pdf (2016).

\bibitem[Park and Newman, 2003]{modularity newman B}Juyong Park and Mark E. J. Newman, ``Origin of degree correlations in the Internet and other networks'', Physical
Review E68 (2003), 026112.

\bibitem{balls and bins paley zigmund}Raymond E. A. C. Paley and Antoni  Zygmund,
``On some series of functions, (3)'',  {Mathematical Proceedings of the Cambridge Philosophical Society}, 28
(1932) 2, 190--205.

\bibitem[Raab and Steger, 1998]{balls and bins Raab n Steger}Martin Raab and Angelika Steger, ``Balls into bins\textendash{}a simple and tight
analysis'', J. D. P. Rolim, M. Serna and M. Luby, Eds., Randomization and Approximation Techniques in Computer Science, 1518, Springer, Berlin (1998), 159\textendash{}170.

\bibitem[Wilson, 1998]{Wilson results}David B. Wilson, http://dbwilson.com/2sat-data/
(1998).

\end{thebibliography}

\newpage
\appendix
\section{Proof of Theorem \ref{thm3.5}}\label{appendixa}
In the proof of Theorem \ref{thm3.5} we shall use the following
lemma, which is analogous to Lemma~\ref{lemma3.4}.
\begin{lem}
\label{lemma 4.1} Consider the balls and bins problem with $B$ bins
and $M(B)$ balls, and also with $B$ bins and~$M^{'}(B)$ balls,
where~$M^{'}(B)=\omega(M(B))$. If the maximum load for~$M(B)$
balls is at least $s\ge1$ with probability bounded away from $0$,
then the maximum load for~$M^{'}(B)$ is at least $s$ w.h.p.
\end{lem}
\begin{Proof}{}
Assume we part the balls into $b(B)=M^{'}(B)/M(B)$ disjoint batches
of $M(B)$ balls each. Suppose we toss the balls in each batch into
the bins separately, and check the maximum load for each batch. Let
$L_{i}$ be the maximum load for batch $i$,~$1\le i\le b(B)$. According
to our assumption, there exists an $\varepsilon>0$ such that 
\[
P\left(L_{i}\ge s\right)\ge\varepsilon,\qquad i=1,2,\ldots,b(B).
\]
Let $L$ be the maximum load in the case we place all the $M^{'}(B)$
balls into the $B$ bins. The events~$\left\{ L_{i}\ge s\right\} ,1\le i\le b(B)$,
are independent, and we clearly have 
\[
\left\{ L<s\right\} \subseteq\bigcap_{i=1}^{b(B)}\left\{ L_{i}<s\right\} .
\]
 Hence:
\begin{align*}
P(L\ge s)=1-P(L<s)\ge1-\left(1-\varepsilon\right)^{b(B)} & \xrightarrow[B\to\infty]{}1.
\end{align*}
 
 \qed
\end{Proof}

The next proof will make use of the notion of negative association
of random variables \citep{Bins and balls dubashi}: Denote~$[k]=\left\{ 1,\ldots,k\right\} $
for $k>0$. Random variables~$X_{1},\ldots,X_{k}$ are $\textit{negatively associated}$
if for every two index sets~$I,J\subseteq[k]$, with $I\cap J=\emptyset$,
\[
E\Bigl(f_{1}(X_{i};i\in I)f_{2}(X_{j};j\in J)\Bigr)\le E\Bigl(f_{1}(X_{i};j\in I)\Bigr)E\Bigl(f_{2}(X_{j};j\in J)\Bigr),
\]
 for every two functions $f_{1}:{\bf R}^{\vert I\vert}\to{\bf R}$
and $f_{2}:{\bf R}^{\vert J\vert}\to{\bf R}$, which are both non-decreasing
or both non-increasing.

In the proof of Theorem \ref{thm3.5}, we will make use of the
following result, concerning the balls and bins problem. Let $Y_{i}$
denote the number of balls placed in the $i$-th bin,~$1\le i\le B$.
Let~$g_{i}:{\bf R\to{\bf R}}$ be non-decreasing functions,~$1\le i\le B$.
By \citep[Lemma 2]{Bins and balls dubashi}, the variables~$Y_{1},\ldots,Y_{B}$
are negatively associated, and in particular the~$g_{i}(Y_{i})$-s
are negatively correlated.

\vspace{10pt}
\begin{Proof}{\ of Theorem \ref{thm3.5}} Let $Y_{1},\ldots,Y_{B}$
be as above. We clearly have~
\[
Y_{i}\sim B\left(M(B),1/B\right),\qquad1\le i\le B.
\]
Define the variables 
\begin{align*}
S_{i} & =
\begin{cases}
1, & \qquad Y_{i}\ge s,\\
0, & \qquad\textmd{otherwise,}
\end{cases}
\qquad 1\le i\le B.
\end{align*}
The $S_{i}$-s are $\textmd{Ber}(p)$-distributed,
where $p=P(Y_{1}\ge s)$. Let $S=\sum_{i=1}^{B}S_{i}$. 

(a) We use the first moment method. Obviously: 
\[
P(S>0)=P(S\ge1)\le E(S)=Bp.
\]
 Let us index the balls from $1$ to $M(B)$, and let $M_{j}=1$ if
the $j$-th ball entered the first bin and $M_{j}=0$ otherwise,~$1\le j\le M(B)$.
Thus, the $M_{j}$-s are $\textmd{Ber}(1/B)$-distributed. Let~$\mathcal{J}=\binom{[M(B)]}{s}$
denote the set of subsets of size $s$ of $[M(B)]$. By the union
bound and symmetry:
\[
\begin{aligned}p=P(Y_{1}\ge s) & =P\Bigl(\bigcup_{J\in\mathcal{J}}\bigcap_{j\in J}\left\{ M_{j}=1\right\} \Bigr)\\
 & \le\dbinom{M(B)}{s}\cdot P\Bigl(M_{1}=\ldots=M_{s}=1\Bigr)=\dbinom{M(B)}{s}\left(\dfrac{1}{B}\right)^{s}.
\end{aligned}
\]
Since $M(B)=o(B^{1-1/s})$,
\[
\begin{aligned}P(S>0) & \le B\cdot\dbinom{M(B)}{s}\left(\dfrac{1}{B}\right)^{s}\le B\cdot\dfrac{M(B)^{s}}{B^{s}}=\left(\dfrac{M(B)}{B^{1-1/s}}\right)^{s}\xrightarrow[B\to\infty]{}0.\end{aligned}
\]
Thus, w.h.p.\ the maximum load does not exceed $s-1$.

(b) We employ the second moment method. First, if $M(B)\geq Bs$, then there must be at least one bin with at least $s$ balls in it. Thus we may assume that $\frac{M(B)}{B}<s$. We have 
\[
\begin{aligned}E(S) & =B\cdot P\left(Y_{1}\ge s\right)=B\cdot\sum_{j=s}^{M(B)}\dbinom{M(B)}{j}\left(\dfrac{1}{B}\right)^{j}\left(1-\dfrac{1}{B}\right)^{M(B)-j}\\
 & \ge B\cdot\dbinom{M(B)}{s}\left(\dfrac{1}{B}\right)^{s}\left(1-\dfrac{1}{B}\right)^{M(B)-s}.
\end{aligned}
\]
 For sufficiently large $B$ we have $M(B)\ge2s$, and therefore 
\begin{align*}
E(S) & \ge B\cdot\dfrac{\left(M(B)/(2B)\right){}^{s}}{s!}\cdot\left(\left(1-\dfrac{1}{B}\right)^{B}\right)^{(M(B)-s)/B}\\
 & \ge B\cdot\dfrac{\left(M(B)/(2B)\right){}^{s}}{s!}\cdot e^{-2M(B)/B}.
\end{align*}
Thus we have
\begin{align}
s! E(S)\ge B\cdot e^{-2M(B)/B}\cdot\left(\dfrac{M(B)}{2B}\right)^{s}\geq e^{-2s}\left(\dfrac{M(B)}{2B^{1-1/s}}\right)^s\label{prop 3.4 (a) label 3}
\end{align}
By \citep{Bins and balls dubashi}, the variables~$Y_{1},\ldots,Y_{B}$
are negatively associated. Since each $S_{i}$ is a non-decreasing
function of $Y_{i}$, this yields~$\textmd{Cov}(S_{i},S_{j})\le0$
for~$i\ne j$. Hence: 
\begin{align*}
V(S) & =\sum_{i=1}^{B}V(S_{i})+2\sum_{1\le i<j\le B}\textmd{Cov}(S_{i},S_{j})\\
 & \le B\cdot V(S_{1})=B\cdot p(1-p)<B\cdot p=E(S).
\end{align*}
 As $S\ge0$, the Paley\textendash Zygmund inequality \citep{balls and bins paley zigmund}
 yields
\begin{equation}
\begin{aligned}P\left(S>0\right) & \ge\dfrac{E^{2}(S)}{E(S^{2})}=\dfrac{E^{2}(S)}{V(S)+E^{2}(S)}>\dfrac{E^{2}(S)}{E(S)+E^{2}(S)}=\dfrac{E(S)}{1+E(S)}.\end{aligned}
\label{prop 3.4 (a) label 4}
\end{equation}
By (\ref{prop 3.4 (a) label 3}), we have $E(S)\xrightarrow[B\to\infty]{}\infty$. Also, by (\ref{prop 3.4 (a) label 3}) and (\ref{prop 3.4 (a) label 4}), we have
\begin{align*}
P\left(S>0\right) & >\dfrac{E(S)}{1+E(S)}
\end{align*}
and so $P\left(S>0\right)\xrightarrow[B\to\infty]{}1$. Thus, w.h.p. the maximum load is at least $s$.

(c) The first statement follows from parts (a) and (b), applied with $s+1$ and $s-1$, respectively, instead of $s$. For the convergence part, suppose \eqref{eqn1} holds. Observe that there are $B^M$ possible ways to distribute the $M$ balls into the $B$ bins. Obviously, $X_B=\sum_{i=1}^B\mathbbm{1}_{y_i=s}$. Let $1\leq t\leq B$. We will prove that 
\begin{equation}\label{eqn2}
\lim_{B\rightarrow\infty}E\binom{X}{t} =\frac{C^{st}}{(s!)^tt!}.
\end{equation}
Specify $t$ bins, say $i_1,i_2,\ldots,i_t$ out of the $B$ bins. The number of balls in bins $i_1,i_2,\ldots,i_t$, and all of the other bins combined forms a multinomial distribution. It follows that
\begin{equation*}
E\binom{X_B}{t}=\frac{\binom{B}{t}\binom{M}{s}\binom{M-s}{s}\binom{M-2s}{s}\cdots\binom{M-(t-1)s}{s}(B-t)^{M-ts}}{B^M}.
\end{equation*}
As $B\rightarrow\infty$, we thus have
\begin{align*}
E\binom{X_B}{t}&=\left(1+o(1)\right)\frac{M^{st}(B-t)^{M-ts}}{t!s!^tB^{M-t}}\\
&=\left(1+o(1)\right)\frac{C^{st}B^{st-t}(B-t)^{M-ts}}{t!s!^tB^{M-t}}\\
&=\left(1+o(1)\right)\frac{C^{st}B^{st}(B-t)^{M-ts}}{t!s!^tB^M}\\
&=\left(1+o(1)\right)\frac{C^{st}(1-t/B)^{-ts}(1-t/B)^M}{t!(s!)^t}.
\end{align*}
From \eqref{eqn1}, we have
\begin{equation*}
\lim_{B\rightarrow\infty}\frac{M}{B}=0
\end{equation*}
and so \eqref{eqn2} holds. The desired result follows from Brun's sieve, which is stated in Theorem $2.1$ of \citep{alon}.

\qed \end{Proof} \bigskip

\end{document}